\documentclass[iop,apj,useAMS,usenatbib]{emulateapj-rtx4}

\usepackage{mystyle}

\begin{document}
\title{Selection of massive evolved galaxies at $3 \leq z \leq 4.5$ in the CANDELS fields}

\author{
Abtin Shahidi\altaffilmark{1}, 
Bahram Mobasher\altaffilmark{1},
Hooshang Nayyeri\altaffilmark{2},
Shoubaneh Hemmati\altaffilmark{3},
Tommy Wiklind\altaffilmark{4},
Nima Chartab\altaffilmark{1},
Mark Dickinson\altaffilmark{5},
Steven L Finkelstein\altaffilmark{6},
Camilla Pacifici\altaffilmark{7},
Casey Papovich\altaffilmark{8},
Henry C. Ferguson\altaffilmark{7},
Adriano Fontana\altaffilmark{9},
Mauro Giavalisco\altaffilmark{10},
Anton Koekemoer\altaffilmark{7},
Jeffery Newman\altaffilmark{11},
Zahra Sattari\altaffilmark{1},
Rachel Somerville\altaffilmark{12, 13}
}
\altaffiltext{1}{University of California, Riverside, 900 University Ave, Riverside, CA 92521, USA}
\altaffiltext{2}{University of California Irvine, Irvine, CA 92697, USA}
\altaffiltext{3}{Jet Propulsion Laboratory, California Institute of Technology, Pasadena, CA, USA}
\altaffiltext{4}{Catholic University of America, Department of Physics, Washington, DC 20064, USA}
\altaffiltext{5}{National Optical Astronomy Observatories, 950 N Cherry Avenue, Tucson, AZ 85719, USA}
\altaffiltext{6}{Department of Astronomy, The University of Texas at Austin, Austin, TX 78712, USA}
\altaffiltext{7}{Space Telescope Science Institute, 3700 San Martin Drive, Baltimore, MD 21218, USA}
\altaffiltext{8}{Department of Physics and Astronomy, Texas A\&M University, College Station, TX 77843-4242, USA}
\altaffiltext{9}{INAF-OAR, Via Frascati 33, Monte Porzio Catone (RM), Italy}
\altaffiltext{10}{Department of Astronomy, University of Massachusetts, 710 North Plesant Street, Amherst, MA 01003, USA}
\altaffiltext{11}{Department of Physics and Astronomy and PITT PACC, University of Pittsburgh, Pittsburgh, PA 15260, USA}
\altaffiltext{12}{Department of Physics and Astronomy, Rutgers University, 136 Frelinghuysen Road, Piscataway, NJ 08854, USA}
\altaffiltext{13}{Center for Computational Astrophysics, Flatiron Institute, 162 5th Ave, New York, NY 10010, USA}
\email{abtin.shahidi@email.ucr.edu}

\begin{abstract}
Using the CANDELS photometric catalogs for the \textit{HST}/ACS and WFC3, we identified massive evolved galaxies at $3 < z < 4.5$, employing three different selection methods. We find the comoving number density of these objects to be $\sim 2 \times 10^{-5}$ and  $8 \times 10^{-6}Mpc^{-3}$ after correction for completeness for two redshift bins centered at $z=3.4, 4.7$. We quantify a measure of how much confidence we should have for each candidate galaxy from different selections and what are the conservative error estimates propagated into our selection. Then we compare the evolution of the corresponding number densities and their stellar mass density with numerical simulations, semi-analytical models, and previous observational estimates, which shows slight tension at higher redshifts as the models tend to underestimate the number and mass densities. By estimating the average halo masses of the candidates ($M_h \approx 4.2,  1.9, 1.3 \times 10^{12} M_\odot$ for redshift bins centered at $z=3.4, 4.1, 4.7$), we find them to be consistent with halos that were efficient in turning baryons to stars and were relatively immune to the feedback effects and on the verge of transition into hot-mode accretion. This can suggest the relative cosmological starvation of the cold gas followed by an overconsumption phase in which the galaxy consumes the available cold gas rapidly as one of the possible drivers for the quenching of the massive evolved population at high redshift.
\end{abstract}
\section{Introduction}
In the standard $\Lambda$CDM paradigm, most of the mass in the universe resides in structures known as dark matter halos. These provide the gravitational well within which cold gas collapses, forms stars, and at a larger scale forms progenitors of the galaxies we observe today. (e.g., \citealt{White1978}; \citealt{Fall1980}; \citealt{Blumenthal1984}; \citealt{Frenk2012}; \citealt{Wechsler2018R}). The dark matter halo itself forms from the gravitational collapse of initial perturbations in the density field at the very early universe (e.g., \citealt{vanAlbada1960}; \citealt{vanAlbada1961}; \citealt{Peebles1970}; \citealt{White1976}). The emergence of the hierarchical model of structure formation (\citealt{Press1974}; \citealt{Gott1975}; \citealt{White1978}) supported by cosmological hydro-dynamical simulations and semi-analytical models suggest a bottom-up scenario, in which massive halos formed from a sequence of mergers (so-called merger trees) and mass accretion as opposed to initial rapid collapse models (e.g., \citealt{White1991}; \citealt{Navarro1991}; \citealt{Katz1992}; \citealt{Kauffmann1993}; \citealt{Lacey1993}; \citealt{Somerville1999}). Discovery of very massive galaxies at high redshifts ($M_s \geq 10^{10} M_\odot$), which constitute most of the luminous baryonic component inside the dark matter halos, however, suggest a rapid build-up of the bulk of their stellar mass at $z>2$,  with intense star formation activity at early times. Observations at submillimeter further confirm the starburst populations with star formation rates exceeding hundreds of solar masses per year (e.g., \citealt{Blain2002}; \citealt{Capak2008}; \citealt{Marchesini2010}; \citealt{Smol2015}). There have been many recent spectroscopic confirmations of such sources at high redshift galaxies experiencing suppressed star formation activity (e.g., \citealt{Whitaker14}; \citealt{Belli2014}; \citealt{Newman2015}; \citealt{Schreiber2018}; \citealt{Newman2018a}; \citealt{Belli2017}; \citealt{Belli2017b}; \citealt{Glazebrook2017}; \citealt{Belli2019}; \citealt{Forrest2020}). To use these systems to constrain galaxy formation and evolution scenarios and to study feedback and quenching mechanisms at early times requires a robust photometric selection of these objects followed by deep spectroscopic observations.

Searching for massive evolved galaxies at high redshifts is challenging due to the faint nature of these galaxies and their small number density, requiring multi-waveband deep imaging data over large areas. The Cosmic Assembly Near-infrared Deep Extragalactic Legacy Survey (CANDELS) (\citealt{Koekemoer2011}; \citealt{Grogin2011}), is a treasury program on the Hubble Space Telescope (HST) providing deep multi-waveband imaging data, allowing detection of such massive systems when the universe was 1-2 Gyrs old. Studying these massive systems, with relatively small star formation activity, can help us understand the mass assembly of galaxies at very early times and estimate the baryonic content of the universe that turned into stars as well as studying primary physical processes responsible for the rapid star formation activity experienced by progenitors of these galaxies. Also, since the universe has a more cold gas reservoir at an early time, we would expect a high level of star-formation that persist longer. Therefore, the low star-formation activity of the massive evolved galaxies requires an explanation about physical mechanisms involved in quenching seen in these galaxies and perhaps other mechanisms for maintaining their low star formation activity. Measuring the evolution of the number/stellar mass density of these systems with relatively high stellar mass (i.e. $M_s \geq 10^{10} M_\odot$) and low specific star formation rate (i.e. sSFR $\leq 10^{-10}$ yr$^{-1}$ for $z \sim 3$ targets) will shed light on some of the outstanding questions regarding early evolution of galaxies.

Over the last two decades, different techniques had been developed to identify different populations of high redshift galaxies (\citealt{Cimatti2002}; \citealt{Roche2002}; \citealt{Daddi2004};  \citealt{Reddy2005}; \citealt{Mobasher2005}; \citealt{Lane2007}; \citealt{Daddi2007}; \citealt{Wiklind2007}; \citealt{Rodighiero2007};  \citealt{Grazian2007}; \citealt{Mancini2009}; \citealt{Fontana2009}; \citealt{Williams2009}; \citealt{Caputi2012};  \citealt{Arnouts2013}, \citealt{Whitaker2013}; \citealt{Barro2014}; \citealt{Nayyeri2014}; \citealt{Straatman2014}; \citealt{Fumagalli2016}; \citealt{Pacifici2016};  \citealt{Siudek2017}; \citealt{Fang2018}; \citealt{Merlin2018}; \citealt{Carnall2018}; \citealt{Carnall2019}; \citealt{Carnall2020}). However, different methods optimized to find the same population of high redshift galaxies, often result in different samples with varying levels of contamination when applied on the same data set.

This paper compares different techniques used for the selection of massive evolved galaxies at high redshifts. Here, we apply these methods to the same data set and compare the results. We then quantify the degree of confidence for each of the detected sources to be a member of the galaxy population in question.

In Section \ref{data}, we present the data. Different selection techniques are introduced in section \ref{selection-methods} and applied on the CANDELS data to identify massive evolved galaxies. In Section \ref{Photometric_errors}, we compare these methods and analyze the effect of photometric errors on each of them. In Section \ref{number-densities}, we compare the number and stellar mass densities of the quiescent galaxies with previous values reported from observations and cosmological simulations. We discuss our results in Section \ref{discussion}. We present our final catalog of massive evolved galaxies in the appendix \ref{appendix}.

We assume a standard cosmology with $H_0=70\,\text{kms}^{-1}\text{Mpc}^{-1}$, $\Omega_b = 0.0486$,  $\Omega_m = 0.3089$, and $\Omega_\Lambda=0.7$ from \cite{Planck2016} unless it is stated otherwise. All magnitudes are in the AB system where $\text{m}_{AB}=23.9-2.5\text{log}(f_{\nu}/1\mu \text{Jy})$ (\citealt{Oke1983}).


\section{Data}\label{data}
We use the latest photometric catalogs from the CANDELS with consistent multi-waveband photometry, and physical parameters for all galaxies to the flux limit of the sample \footnote{\hyperlink{http://arcoiris.ucolick.org/candels/}{http://arcoiris.ucolick.org/candels/}}. Details about the selection and photometry at different bands were carried out for all the CANDELS fields consisting of: GOODS-South (\citealt{Guo2013}; \citealt{Santini2015}), UDS (\citealt{Galametz2013};\citealt{Santini2015}) , COSMOS (\citealt{Nayyeri2017b}) , EGS (\citealt{Stefanon2017}), and GOODS-North (\citealt{Barro2019}). These catalogs contain observed photometry from the UV to near and mid-infrared wavelengths in many broadband and narrowband filters (Table \ref{table:filtersets}), as well as inferred physical parameters.

\setlength{\arrayrulewidth}{0.25mm}
\setlength{\tabcolsep}{6pt}
\renewcommand{\arraystretch}{1.3}

\begin{table*}[ht!]
\centering
    \caption{The observed bands from UV-to-NIR SEDs of galaxies across the five CANDELS fields}

    \begin{tabular}{c|c}
    \hline 
    \textbf{Field}  & \textbf{Filter set} \\
    \hline \\

    GOODS-S (\citealt{Guo2013}) & Blanco/CTIO U, VLT/VIMOS U, \\
    & HST/ACS F435W, F606W, F775W, F814W, F850LP, \\
    & HST/WFC3 F098M, F105W, F125W, F160W, \\
    & VLT/ISAAC Ks, VLT/Hawk-I $K_s$, \\
    & Spitzer/IRAC 3.6 $\mu$m,  4.5 $\mu$m,  5.8 $\mu$m,  8.0 $\mu$m   \\

    \\
    
    \hline 
    
    \\

    GOODS-N (\citealt{Barro2019}) & KPNO U, LBC U, \\
    & HST/ACS F435W, F606W, F775W, F814W, F850LP, \\
    & HST/WFC3 F105W, F125W, F140W, F160W, F275W, \\
    & MOIRCS K, CFHT $K_s$, \\
    & Spitzer/IRAC 3.6 $\mu$m,  4.5 $\mu$m,  5.8 $\mu$m,  8.0 $\mu$m \\
    
    \\
    
    \hline 
    
    \\
    
    UDS (\citealt{Galametz2013}) & CFHT/MegaCam u, Subaru/Suprime-Cam B, V, $R_c$, $i^{\prime}$ , $z^{\prime}$ , \\
    & HST/ACS F606W, F814W, HST/WFC3 F125W, F160W, \\
    & VLT/Hawk-I Y, $K_s$, \\
    & WFCAM/UKIRT J, H, K, \\
    & Spitzer/IRAC 3.6 $\mu$m,  4.5 $\mu$m,  5.8 $\mu$m,  8.0 $\mu$m \\

    \\
    
    \hline 
    
    \\

    EGS (\citealt{Stefanon2017}) & CFHT/MegaCam $U^*$, $g^{\prime}$ , $r^{\prime}$ , $i^{\prime}$ , $z^{\prime}$ , \\
    & HST/ACS F606W, F814W, HST/WFC3 F125W, F140W, F160W, \\
    & Mayall/NEWFIRM J1, J2, J3, H1, H2, K, \\
    & CFHT/WIRCAM J, H, $K_s$,  \\
    & Spitzer/IRAC 3.6 $\mu$m,  4.5 $\mu$m,  5.8 $\mu$m,  8.0 $\mu$m \\

    \\
    
    \hline  
    
    \\

    COSMOS (\citealt{Nayyeri2017b})  & CFHT/MegaCam $u^{*}$, $g^*$, $r^*$, $i^*$, $z^*$, \\
    & Subaru/Suprime-Cam B $g^+$, V, $r^+$, $i^+$, $z^+$, \\
    & HST/ACS F606W, F814W, HST/WFC3 F125W, F160W, \\
    & Subaru/Suprime-cam IA484, IA527, IA624, IA679, IA738, IA767, IB427,  \\
    & IB464, IB505, IB574, IB709, IB827, NB711, NB816, \\
    & VLT/VISTA Y, J, H, $K_s$, Mayall/NEWFIRM J1, J2, J3, H1, H2, K,  \\
    & Spitzer/IRAC 3.6 $\mu$m,  4.5 $\mu$m,  5.8 $\mu$m,  8.0 $\mu$m \\
    
    \\
    \hline
\end{tabular}
    \label{table:filtersets}
\end{table*}
The optical (\textit{HST}/ACS) and near-IR (\textit{HST}/WFC3) data  were consistently combined with the mid-IR data (\textit{Spitzer}/IRAC) and ground-based observations.
For each of the CANDELS fields, the photometric catalogs were selected in \textit{HST}/WFC3 F160W band using SExtractor (\citealt{Bertin1996}). 
For low-resolution images, Template FITting (\textsc{Tfit}; \cite{Laidler2007}) was performed to smooth the high-resolution image to low resolution and fit the best flux consistent with the \textit{HST} data. \textsc{Tfit} creates a template using prior information on the position and light distribution profile of sources in the high-resolution to more robustly measure photometries from the lower resolution data. Table \ref{table:CANDELS_area} lists the limiting magnitudes and survey areas covered for each of the CANDELS fields. 

\begin{table*}[t!]
\centering
\begin{tabular}{ccc}
\hline 
 \textbf{Field} & \textbf{Area ($arcmin^2$)} & \textbf{F160W $5\sigma$ Limiting Magnitude (AB)} \\
\hline 
GOODS-North & 170 & 27.80 \\
GOODS-South & 170 & 27.36\\ 
COSMOS & 216 & 27.56 \\
UDS & 202 & 27.45 \\
EGS & 206 & 27.60 \\
\hline
\end{tabular}
\caption{The survey area and WFC3 F160W limiting magnitude for different CANDELS fields}
    \label{table:CANDELS_area}
\end{table*}

In this study, we use an improved version of the original \citealt{Dahlen2013} photometric redshifts. The new catalogs are based on the probability density functions (PDF) measured by six groups using different template-based methods applied on the CANDELS photometric catalogs. The methods are different either in their choice of parameters or the code used. The PDFs from different groups were corrected and optimized for bias (optimal shift) and their variance (optimal scaling of the width of the PDF). After re-calibration, the PDFs were combined based on the minimum Frechet distance \cite{ALT1995}, which tracks the similarity between any pair of PDF curves, among the independent photo-z PDFs (analogous to the median of a set of numbers). The final catalog consist of the spectroscopic/3D-\textit{HST} grism redshifts and the combined photometric redshifts (Kodra et al. in prep.). The point estimate redshifts used in this work show a normalized median absolute deviation of $\sigma_{\text{NMAD}} \sim 0.02$. The stellar masses were measured through SED fitting by keeping redshifts at their best values.

\section{Selection Methods} \label{selection-methods}
Here we use a combination of several near-infrared selection techniques to identify massive quiescent galaxies at a redshift of $3 \leq z\leq 4.5$ in the CANDELS fields. This is the redshift interval that most of the galaxies exist; however, the full sample of galaxies is at a redshift of $2.8 \leq z \leq 5.4$. We define the massive quiescent population as those with $M_s \geq 10^{10} M_\odot$ and $sSFR(z)\leq 0.2/t_H(z)$ Gy$^{-1}$ in which $t_H(z)$ is the age of the universe at redshift $z$ in Gyrs. The results are cross-compared and used to compile a reliable catalog for the study of the number density and mass function of these systems in the redshift range mentioned. This comparison also allows a better understanding of the strengths and shortcomings of each method.

First, we use the color-selection technique based on rest-frame or observed colors for classification. Then, we explore selection methods based on the inferred physical properties of galaxies from their best-fitted model Spectral Energy Distributions (SEDs).

A serious source of uncertainty in these techniques is the choice of the somewhat subjective parameters that could significantly affect the outcome. These include the S/N in the flux in a particular band, color cuts in the color-space occupied by galaxies, flux limits, and parameters used to generate template SEDs. Galaxies that are close to the decision boundaries are most likely to be affected by the choice of these parameters. Therefore, to minimize the effect of these cuts for each technique, we define a likelihood function between zero and one that identifies the likelihood a given galaxy is a real member of the massive quiescent population within our specified redshift range. We then consistently combine all these measures to find a final value associated with individual galaxies to quantify the degree of confidence for each selection method and to compare different galaxies resulted from a particular selection. Then we estimate the uncertainty in number/stellar mass density measurements coming from the selection thresholds. We refer the reader to the Appendix for more details. Following this procedure, we select $5-10\%$ more candidates than using a step-function selection that misses galaxies close to the selection boundary. This difference is significant enough to affect the measurement of their number and mass densities. In the following section, we describe each selection method.

\subsection{Rest-frame UVJ selection} \label{uvj}
This method uses rest-frame U, V, and J bands to select the quiescent population, as the rest frame $U-V$ probes the prominent Balmer break ($3646\AA$) seen in post-starbursts. The $V-J$ color used to break the degeneracy between dusty star-forming and quiescent galaxies. We can classify galaxies using a color cut that separates the quiescent and star-forming regions in the UVJ plane. This method has been developed and extensively used for classifying galaxies in photometric surveys (e.g.,  \citealt{Labbe2005}; \citealt{Wuyts2007}; \citealt{Williams2009}; \citealt{Arnouts2013}, \citealt{Whitaker2013}; \citealt{Barro2014}; \citealt{Straatman2014}; \citealt{Fumagalli2016}, \citealt{Siudek2017} \citealt{Fang2018}). However, measuring UVJ colors for galaxies at high redshifts is challenging, making the selection less reliable. A typical method to infer rest-frame UVJ colors is from the best-fitted model SED. The model SEDs are generally built assuming a $\tau$ model star formation history, which also becomes less reliable at high redshifts since the galaxy model needs more time to evolve into the quiescent regions (\citealt{Merlin2018}) and more generally suffer from other uncertainties associated with SED fits used to infer rest-frame colors. Therefore, the boundaries of the UVJ criteria are modified at different redshift bins (\citealt{Whitaker2013}). However, this can introduce contamination from the dusty star-forming galaxies, given the similarity of their rest-frame $U-V$ color. Nevertheless, given the overlap in the wavelength coverage, including medium-band data, can ameliorate some of these problems when selecting high-z galaxies (e.g., \citealt{Spitler2014}).

To measure the rest-frame UVJ colors, we employ the prescription described by \citealt{Wolf2003} for the COMBO-17 survey, where the rest-frame colors are estimated from the best fit template SED. Rest-frame UVJ colors can also be measured by interpolating the observed bands which track the redshifted UVJ (e.g., \citealt{Rudnick2003}, \citealt{Taylor2009}, \citealt{Williams2009}). In the latter case, data from many bands outside of the UVJ region of the spectra is left unused while in constraining the best fit model SED, these data are beneficial (\citealt{Brammer2011}). We will further investigate the effect of the star-formation histories on the UVJ colors from best-fit SEDs when analyzing the effects of the photometric uncertainties in Section \ref{Photometric_errors}.

\begin{figure}
    \centering
    \includegraphics[scale=0.4]{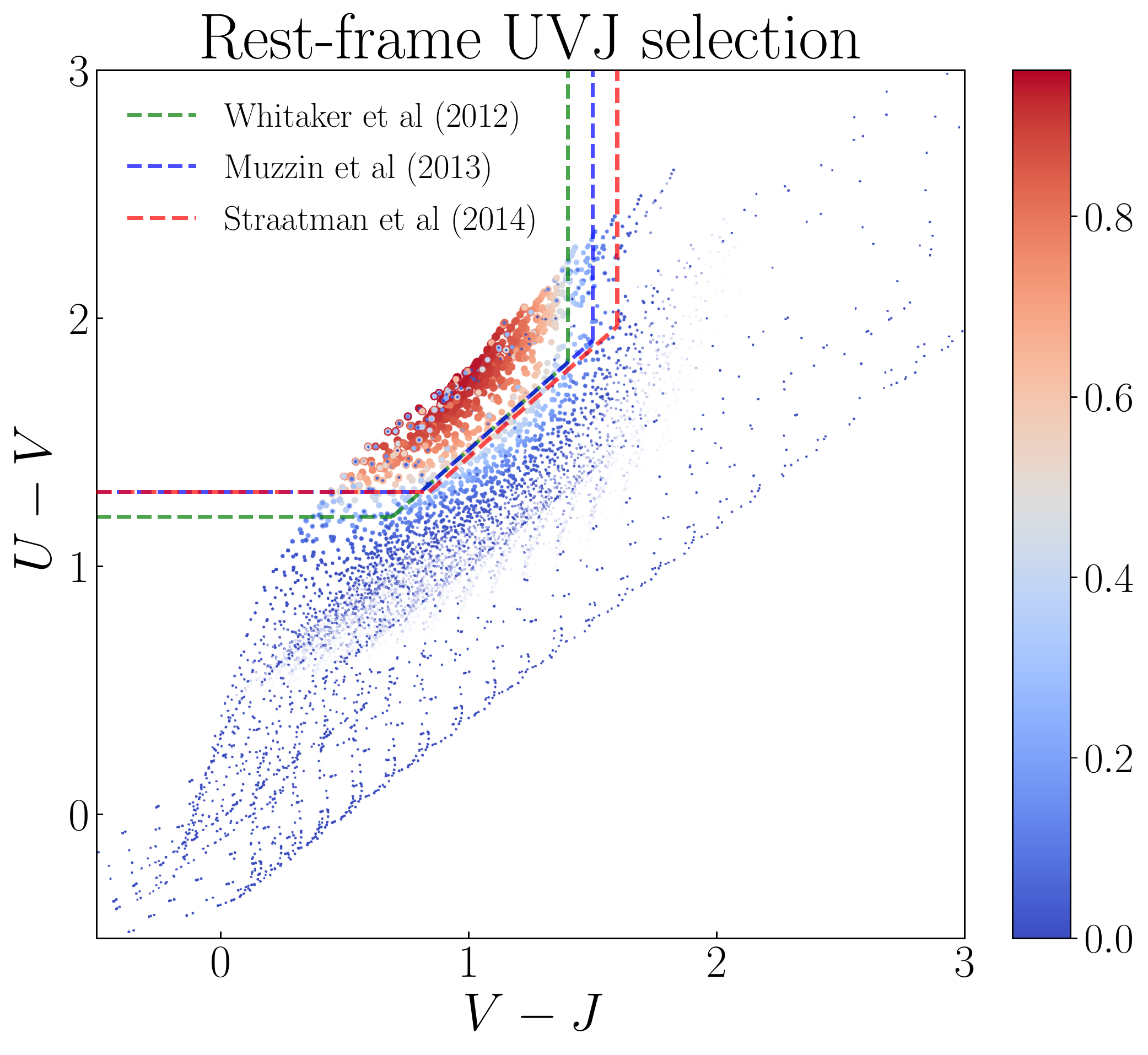}
    \caption{The rest-frame UVJ plane and colors from the best-fitted models for every galaxy are shown. The artificial patterns seen are from the limited number of models that do not span the total dynamic range of the UVJ colors of galaxies. Also, since we have imposed another criterion on the F125W-band, we do not select the low-z dusty solution counterparts. The color bar shows the likelihood grading for each selection. Redder data points are more likely to be quiescent. The likelihoods are assigned based on the prescription defined in Appendix \ref{appendix}.}
    \label{fig:uvj_selection}
\end{figure}

We combine the selection criteria on the rest-frame $UVJ$ plane employed in \citealt{Muzzin2013}, \citealt{Whitaker2011}, and \citealt{Straatman2014} (Figure \ref{fig:uvj_selection}). We define the grading scheme by giving a weight to individual galaxies based on their distance from the boundary defined in Table \ref{table:selections}. (Figure \ref{fig:uvj_selection}). The combination of these selections is done by finding the likelihood that each galaxy is a member of the union of selected populations from different criteria. To find the rest-frame UVJ colors of CANDELS galaxies, we use the \textsc{Le Phare} SED fitting code (\citealt{Arnouts1999}, \citealt{Ilbert2006}) with standard libraries defined in Table \ref{table:SED} (\citealt{Chartab2020}).


\begin{table*}[t!]
\centering
\begin{tabular}{c|c|c|c|c|c}
\hline
\textbf{SPS model} & \textbf{$\tau$ Gyr}& \textbf{$E(B-V)$} & \textbf{IMF} & \textbf{Dust attenuation law} & \textbf{Metallicity} \\
\hline 
BC03 (\cite{Bruzual2003}) &  (0.01, 30)  & (0, 1.1) & \cite{Chabrier2003} & \cite{Calzetti2000} & \{0.02, 0.008, 0.004\}\\

\hline

\end{tabular}
\caption{The range of the parameter used in the \textsc{Le Phare} code for exponentially declining star formation history, as well as the initial mass function and dust attenuation law used for fitting all the CANDELS.}
\label{table:SED}
\end{table*}

In addition to the criteria on the UVJ plane, we impose a mild detection constraint on the observed J-band ($S/N>2$), which probes blueward of the Balmer break for the highest redshift bins. This reassures that the break lies redward of the J-band, reducing the contamination from low-z interlopers. The results are presented in Figure \ref{fig:uvj_selection}. 


\subsection{Observed-color selection} \label{observed-color}
The most commonly used methods for identifying the population of high redshift galaxies are variations of the drop-out technique, based on the observed colors of galaxies. This uses well-known features in the galaxy SEDs and follows them as they move to redder passbands when galaxy is redshifted. Examples of this are the Lyman break features used for the selection of UV bright Lyman break Galaxies (\citealt{Steidel1993}; \citealt{Steidel1995}; \citealt{Steidel2003};  \citealt{Reddy2005}; more recently \citealt{Stark2010}; \citealt{Bouwens2011}; \citealt{Bouwens2014}; \citealt{Roberts2016}; \citealt{Oesch2016}) and evolved systems using Balmer break features (\citealt{Cimatti2002}; \citealt{Roche2002}; \citealt{Franx2003}; \citealt{vanDokkum2003}; \citealt{Daddi2004};  \citealt{Reddy2005}; \citealt{Mobasher2005}; \citealt{Lane2007}; \citealt{Daddi2007}; \citealt{Wiklind2007}; \citealt{Rodighiero2007}; \citealt{Caputi2012};  \citealt{Nayyeri2014}; \citealt{Girelli2019}).

This technique uses the fact that magnitudes and colors are sensitive to redshift and the shape of their SEDs, which follows the physical properties of their stellar population and their interstellar medium.  For example, for post-starburst galaxies, we can use Balmer break features from $3648 \AA$ Balmer limit. Therefore, the observed colors can help us find the candidate galaxies directly from photometric measurements using a few photometric bands. For the redshift range of interest here, Balmer break features redshift towards near-IR wavelengths.

\begin{figure}
    \includegraphics[scale=0.35]{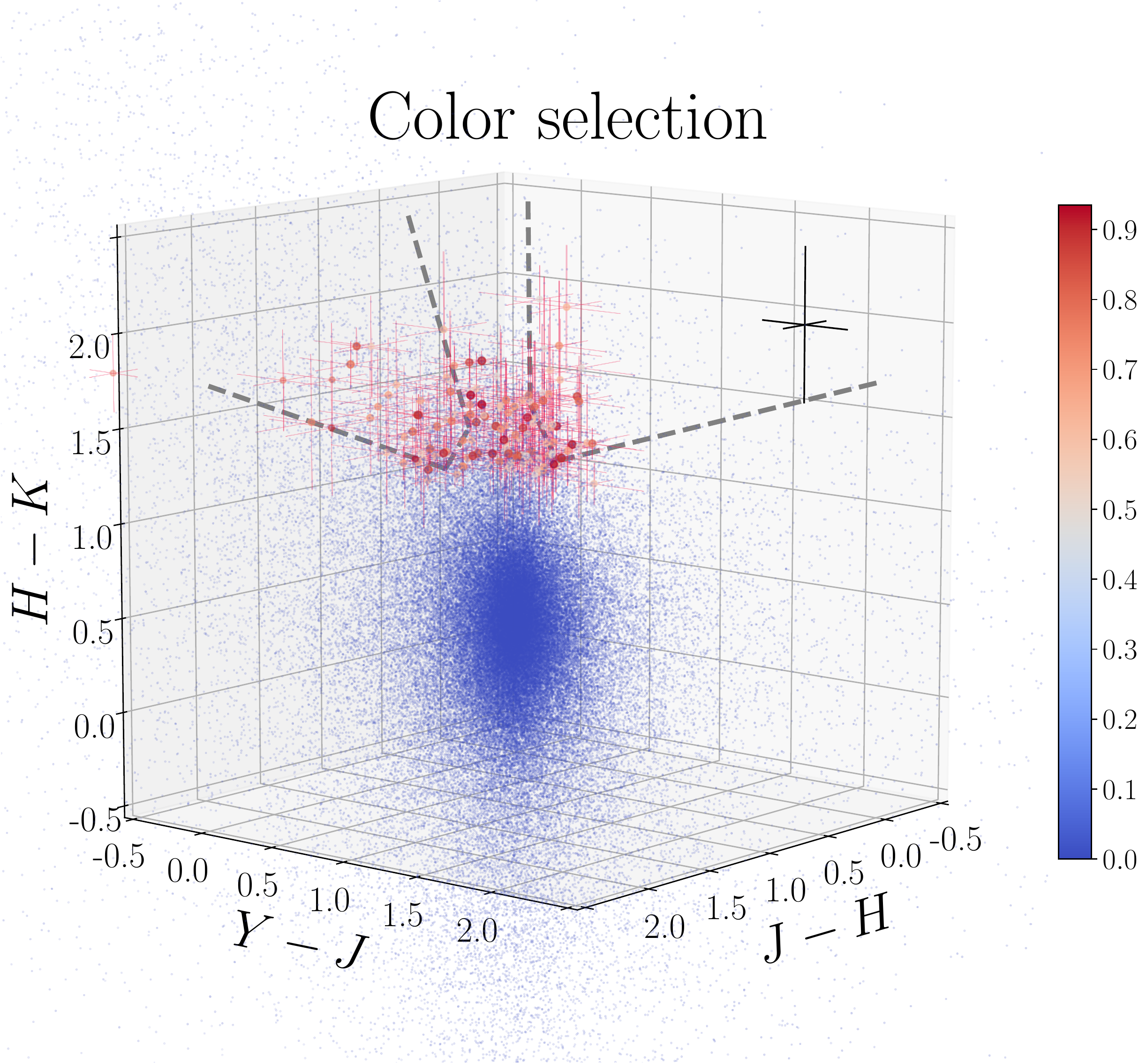}
  \caption{Observed color selection based on the criteria in \citealt{Nayyeri2014} in 3 dimensions is shown (The gray dashed line shows these selection boundaries on the 2D planes of the colors). The redness of the points indicates the degree to which each galaxy is a member of the quiescent population. The error bars are plotted for sources with a likelihood higher than 0.5. The likelihoods are assigned based on the prescription defined in Appendix \ref{appendix}.}
    \label{fig:bbg_selection3d}
\end{figure}

We use the observed colors to select candidates in the CANDELS fields with the color selection criteria from \citealt{Nayyeri2014} (Figure \ref{fig:bbg_selection3d}). This selection was initially developed by finding the color criteria in the color-color space for a sample of old (low-z) quiescent, dusty starburst, and post starburst populations, using \citealt{Bruzual2003} stellar population synthesis evolutionary tracks and considering an IGM absorption following \citealt{Madau1995}. In the redshift range of interest ($3 \leq z \leq 4.5 $), $H-K$ color will constrain the presence of the Balmer break, accompanied by a $Y-J$ or $J-H$ constraint which can discriminate between dusty-starburst and quiescent galaxies. Also, we impose a non-detection requirement on the U and B bands since they fall blueward of the break. By doing this, we reduce the contamination from the low redshift interlopers. Another likely source of contamination is due to the absence of nebular emission lines. These affect the selection based on broadband photometry since they can mimic a Balmer break feature, which can mislead the classification scheme (see \citealt{Nayyeri2014}; \citealt{Straatman2014}; \citealt{Merlin2018}). To minimize the effect of these lines, we perform the SED-fitting procedure described in Section \ref{sed-fitting} using libraries with and without the nebular line emissions.

We assign likelihoods to the selected sources, based on their proximity to the selection boundaries to reduce the dependence on the chosen selection boundaries. (More details are presented in the Appendix) The final likelihood will be the average likelihood for all of the realizations of a galaxy within its error bars (Section \ref{errors-observed}).

\subsection{Selection based on SED fitting} \label{sed-fitting}
Here we discuss another method used for finding quiescent candidates based on the inferred physical properties measured from their SEDs. The advantage of this method is that it makes use of all the photometric data available and, therefore, imposes stronger constraints on the selection process. Furthermore, it predicts the physical parameters for each galaxy. The disadvantage is that the method is model dependent and is based on the star formation history and extinction used to generate SED templates (\citealt{Grazian2007}; \citealt{Fontana2009}; \citealt{Pacifici2016}; \citealt{Merlin2018}; \citealt{Carnall2018}; \citealt{Carnall2019}; \citealt{Merlin2019}).  We fit template SEDs for the candidates between $2.8<z_{phot}<5.4$ with masses larger than $10^{10} M_\odot$ using CANDELS photometric and spectroscopic redshifts (where available), and catalog of the physical properties. We rely only on the mass measurements since the stellar masses are less susceptible to the parameters chosen in the template libraries of galaxies, particularly the star formation history used for SED fitting (compare to the inferred SFRs)(\citealt{Mobasher2015}; \citealt{Papovich2001}; \citealt{Shapley2001}; \citealt{Wuyts2009}; \citealt{Muzzin2009}). However, the presence of the nebular lines can affect stellar mass measurements. Hence, we treat the stellar mass as a free parameter (although we made a prior mass selection on the sub-sample) when fitting the sub-samples photometric measurements to keep the stellar mass uncertainty of this type limited to the initial selection.

We find the physical properties using Bayesian Analysis of Galaxies for Physical Inference and Parameter EStimation (\textsc{Bagpipes}), a Bayesian SED fitting code (see \citealt{Carnall2018}). \textsc{Bagpipes} uses 2016 version of the \citealt{Bruzual2003}, and \textsc{Multinest}  (\citealt{Feroz2008}; \citealt{Feroz2009}) for multimodal nested sampling algorithm (\citealt{Skilling2006}). The multimodal nested sampling algorithm employed is a huge improvement over a simple $\chi^2$ fit, which is incapable of producing a reliable error estimate. Moreover, the Markov chain Monte Carlo (MCMC) algorithm employed in some SED-fitting procedures for finding the posterior of the physical properties can be problematic when sampling from multimodal posterior. These include models with large degeneracy, which is often the case when modeling a complex system such as a galaxy under large photometric uncertainties.

We fixed redshifts to their photo-z, and where available, the spec-z values. We built separate model libraries based on three prescriptions for  SFHs: exponentially declining, top-hat, and double power-law forms (\citealt{Behroozi2013}), assuming \citealt{Chabrier2003} IMF,  dust attenuation based on \citealt{Calzetti2000}, and IGM absorption from \cite{Inoue2014} which is a revised version of the \cite{Madau1995} model.  Knowing that nebular lines emission can mimic a Balmer break-like feature and for controlling their effects on the physical properties and, consequently, the selection made based on them, we run the code with and without the nebular emission (since the target population is expected to have little to no nebular emissions). The priors used for physical parameters in the SED fitting are listed in Table \ref{table:priors}. We then find the posterior distribution for each model parameter. Following \cite{Carnall2018}, we define $\psi_{SFR}$ as the ratio of the $SFR$ at any given time ($SFR(t)$) to the average $SFR$ over the age of a given galaxy ($<SFR(t)>$):


\begin{table*}[t!]
\centering
\begin{threeparttable}[t]
\caption{These are the priors used for different free physical parameters and the fixed parameters used in the fit.}
\begin{tabular}{c|ccc|cc}
\hline
\textbf{Star Formation History} & \textbf{Free Parameter} & \textbf{Prior} & \textbf{Limits}  &  \textbf{Fixed Parameter} & \textbf{Value} \\

\hline
& $A_V$\tnote{1} & Uniform & (0, 2)  & SPS models & BC03 \\

Double Power Law: & $\mathrm{log_{10}}(M_\mathrm{formed}\ /\ \mathrm{M_\odot})$\tnote{2}  & Uniform & (1, 13) &   IMF & \cite{Chabrier2003}\\

& $\mathcal{Z}\ /\ \mathcal{Z}_\odot$\tnote{3} & Uniform & (0.2, 2.5) &   $z_\mathrm{obs}$ & $z_\mathrm{phot/spec}$  \\

$SFR(t) = C [(t/\tau)^\alpha + (t/\tau)^{-\beta} ]^{-1}$ & $\tau \ /\ \mathrm{Gyr}$\tnote{4} & Uniform & (0, $t(z_\mathrm{obs}))$ &  $\log_{10}(U)$\tnote{5} & -3\\

& $\alpha$\tnote{6} & Logarithmic & ($10^{-2}, 10^{3}$)  &  \\

& $\beta$\tnote{7} & Logarithmic & ($10^{-2}, 10^{3}$)   &  \\

\hline
 & $A_V$  & Uniform & (0, 2) &  SPS models & BC03\tnote{12}\\

Exponentially Declining: & $\mathrm{log_{10}}(M_\mathrm{formed}\ /\ \mathrm{M_\odot})$  & Uniform & (1, 13) &   IMF & \cite{Chabrier2003}\\

& $\mathcal{Z}\ /\ \mathcal{Z}_\odot$ & Uniform & (0.2, 2.5) &    $z_\mathrm{obs}$ & $z_\mathrm{phot/spec}$  \\
$SFR(t) = C \exp{(-t/\tau)}$ & $\tau \ /\ \mathrm{Gyr}$\tnote{8} & Uniform & (0.05, 10)  & $\log_{10}(U)$ & -3\\

& Age & Uniform & (0, $t(z_\mathrm{obs}))$ &  \\

\hline
& $A_V$  & Uniform & (0, 2) &  SPS models & BC03\\

Top-Hat: & $\mathrm{log_{10}}(M_\mathrm{formed}\ /\ \mathrm{M_\odot})$  & Uniform & (1, 13) &  IMF & \cite{Chabrier2003}\\

 & $\mathcal{Z}\ /\ \mathcal{Z}_\odot$ & Uniform & (0.2, 2.5) &  $z_\mathrm{obs}$ & $z_\mathrm{phot/spec}$ \\
$SFR(t) = \begin{cases} C & t \in [\text{Age}_\text{min},\text{Age}_\text{max}]\\
       0 &\text{otherwise}\end{cases}$ & Age$_\text{min}$\tnote{9} & Uniform & (0, $t(z_\mathrm{obs})) $ &   $\log_{10}(U)$ & -3\\

& Age$_\text{max}$\tnote{10} & Uniform & (0, $t(z_\mathrm{obs})$) & \\
\hline
\end{tabular}
\label{table:priors}
\begin{tablenotes}\footnotesize
  \item [1] $A_v$ is the  Attenuation at $5500$\AA
  \item [2] $M_{formed}$ is the mass formed
  \item [3] $\mathcal{Z}$ is the metallicity
  \item [4] $\tau$ is the peak time for double power law SFH
  \item [5] $\log_{10}U$ is the ionization parameter
  \item [6] $\alpha$ is the rising power
  \item [7] $\beta$ is the falling power
  \item [8] $\tau$ is the exponential decay timescale in $\tau$ model SFH
  \item [9] Age$_{min}$ is the initial time for Top-hat SFH
  \item [10] Age$_{max}$ is the final time for the Top-hat SFH
  \item [11] $C$ is the normalization constant
  \item [12] BC03 is the Stellar population synthesis at the resolution of 2003 (\citealt{BC03})
\end{tablenotes}
\end{threeparttable}
\end{table*}

 \begin{equation}
    \psi_{SFR} = \frac{SFR(t)}{<SFR>(t)} = \frac{SFR(t)}{\frac{1}{t} \int_0^t SFR(t') dt'}
 \end{equation}

 \noindent and we define the quiescent population as those with $\psi_{SFR}$ less than 0.1 at the observed age of the galaxy. In other words, the quiescent galaxies are those with average $SFR$ over the last $100$ Myrs to be less than 10 percent of the average $SFR$ over its lifetime.


\citealt{Carnall2018} showed that the selection criterion mentioned above is consistent with the definition proposed in \citealt{Pacifici2016}, (Figure \ref{fig:comparison}), which is a criterion on the specific star formation rate evolving with redshift to define the quiescent population at a given age of the universe.

The Bayesian nature of the SED fitting code allows us to have the posterior distribution for each galaxy and all the parameters associated with it. Then we can apply the selection on the sample from the posterior and count the number of selected samples from the posterior versus the total count. We assign this ratio as the probability of being selected given the posterior sample. We apply the same prescription for the rest of the models employed.

\begin{figure}
    \includegraphics[scale=0.26]{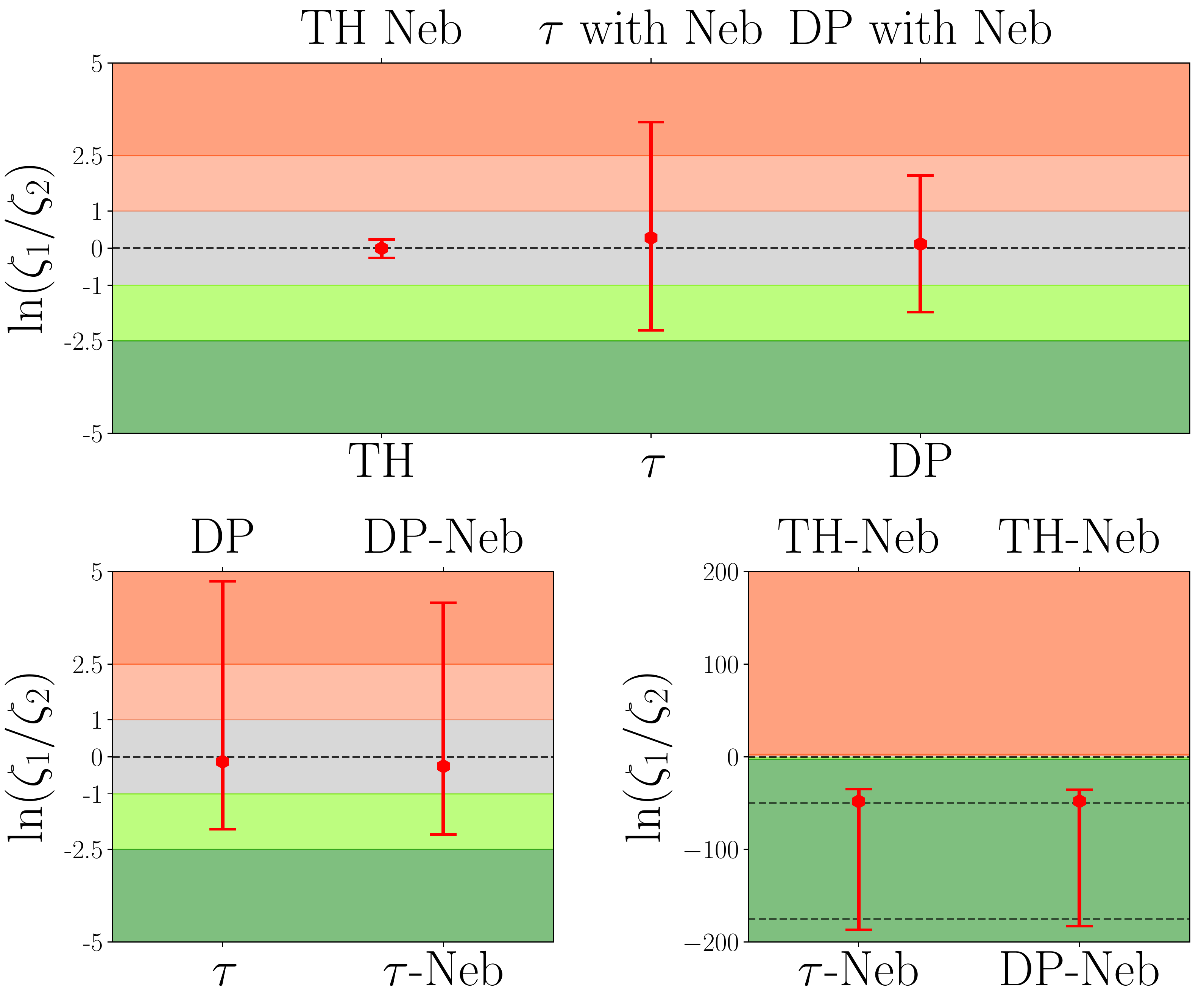}
  \caption{Shows the Bayes factor for models with different SFHs and with/without nebular emissions for the redshift and mass-selected sample of galaxies. TH, $\tau$, and DP stand for Top-Hat, exponentially declining, and double power-law model for SFH, respectively. The top panel shows the Bayes factor for models that include nebular emission to those with the same SFH but without emissions. The bottom left shows the relative evidence for Top-hat SFH to $\tau$ models and double power law. The bottom right shows the relative evidence for $\tau$ models and double power law. The data points show the median, and the error bars are 20, and 80 percentile of the sample. $\zeta_1$ and $\zeta_2$ are evidence for top and bottom models (in each figure). The grey area shows Inconclusive to Weak evidence, the lighter shade shows Weak to Moderate evidence, and darker shade shows Moderate to Strong evidence (\citealt{Jeffreys61}). In the case of top hat models, there is strong evidence against the top-hat models when compared with $ \tau $ and double power law.  }
    \label{fig:evidence}
\end{figure}

We use the Bayesian evidence (marginal likelihood) on different models used for SED-fitting. By doing so, we find the relative evidence for different models according to the data. The definition of the Bayesian evidence is:

\begin{equation}
         \zeta = p(D|H) = \int p(D|\theta,H) p (\theta|H) d\theta 
\end{equation}

\noindent where $D$, $H$, and $\theta$ represent the data, hypothesis (model), and model parameters, respectively. The $p(D|H)$ is the probability of getting the data $D$ under the assumption of the validity of a specific hypothesis $H$, which is the Bayesian evidence for $H$. 

We adopt Jefferey's criteria (\citealt{Jeffreys61}; also see \citealt{Kass1995a}) for interpretation of the relative evidence (Bayes factor) calculated for each galaxy and different models. We use the same prior for fixed parameters such as redshift, and for the rest of the parameters, we use non-informative priors (\citealt{Kass1995b}) across different models. By doing so, the Bayes factor represents the posterior odds of the models ($\frac{p(H_2|D)}{p(H_1|D)}= \frac{p(D|H_1)}{p(D|H_2)}$). In other words, we impose no prior judgment about the validity of a certain model. Results presented in Figure (\ref{fig:evidence}) express that statistically, none of the models show substantially more evidence. However, the Bayes factors support models that include nebular lines and generally favor double power-law over $\tau$-models. This is consistent with \citealt{Reddy2012b} regarding the inability of the $\tau$ models in reproducing star formation rates from UV+IR measurements in the star-forming population at higher redshifts. \citealt{Pacifici2016} showed that double power-law is the best model to describe the SFHs of their samples. \citealt{Carnall2019b} also confirmed that a double power-law model could produce relatively stronger evidence. As Figure \ref{fig:evidence} depicts, similar to \citealt{Belli2019} investigation of the effect of SFH on age measurements, top-hat models generally fail to produce a comparable level of evidence from data. Therefore, we do not use the results from the Top-Hat models for making a selection.

\subsection{AGN Contamination}
We cross-matched the potential candidates for high redshift massive evolved systems with the publically available Chandra X-ray catalogs and used Spitzer MIPS 24 $\mu$m to detect possible dusty AGNs. We excluded any candidates with a counterpart (shows detection) in either X-ray or MIPS. For the X-ray catalogs we looked for any counterparts closer than 5 arcsec which is 10 times the resolution of the Chandra X-ray Observatory (cross-matched with \citealt{Laird2009}; \citealt{Evans2010}; \citealt{Xue2016}; \citealt{Nandra2015};  \citealt{Cappelluti2016}; \citealt{Civano2016}; \citealt{Massini2018}) and for Mid-IR we used the catalog published in \citealt{Barro2019} for finding any counterpart and flux measurements in the mentioned MIR bands. About $30\%$ of the candidates have an X-ray or MIR counterpart, which were removed from the final sample.

\subsection{Final sample}
\begin{figure*}
    \hspace{-0.55cm}
    \includegraphics[scale=0.27]{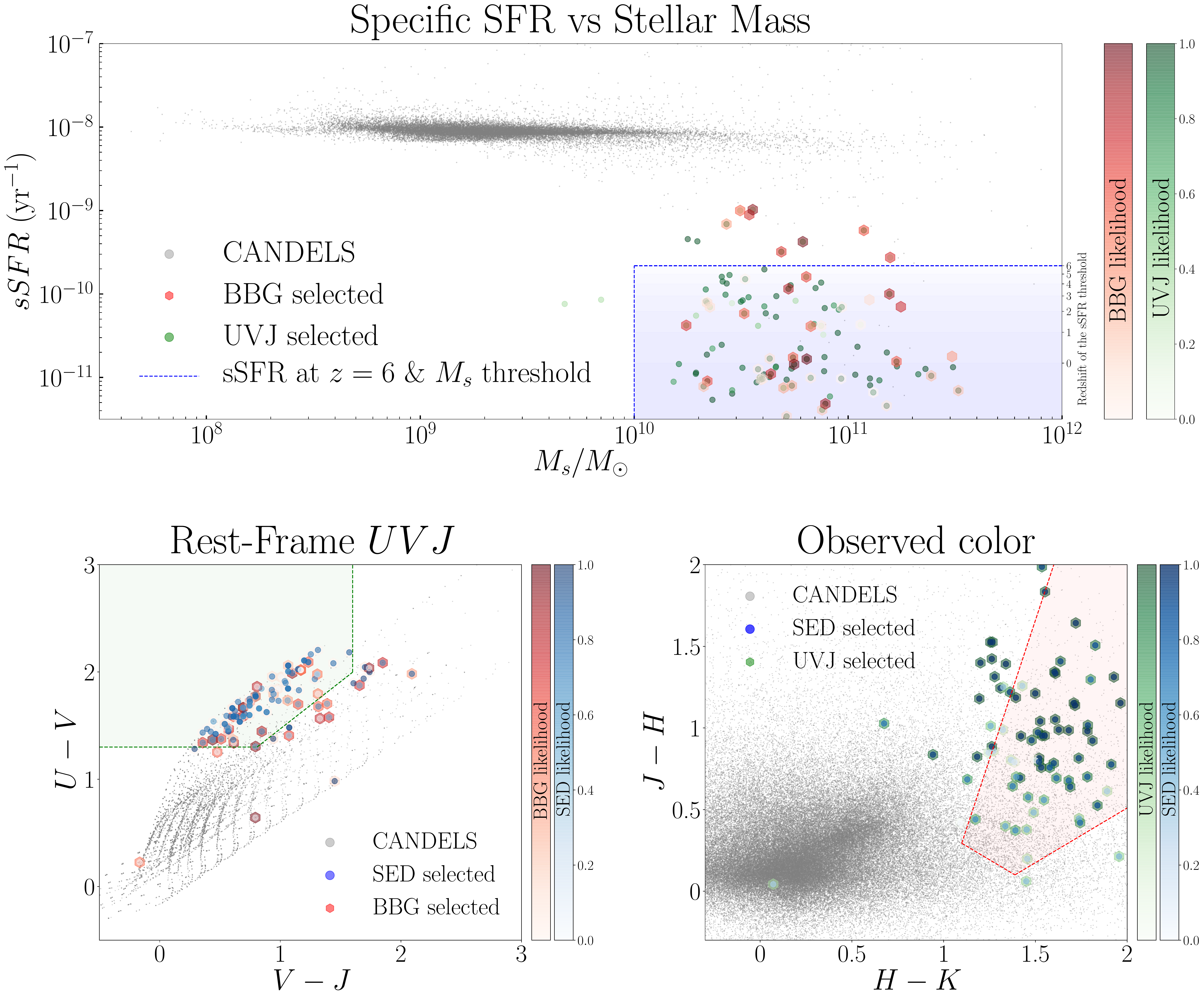}
  \caption{ The top panel shows the $sSFR$ vs $M_s$, (from \textsc{Bagpipes}) the bottom left shows the rest-frame UVJ colors (from \textsc{LePhare}), and the bottom right shows the observed colors for the candidate galaxies. The grey points are the CANDELS galaxies (with $2.8 \leq z \leq 5.4$), and the candidates are color-coded with three assigned likelihoods. The green, blue, and red color scales reflect the likelihood measured based on UVJ, SED, and the observed colors, respectively. In order to show the contrast in different selections, we plotted likelihoods on top of each other. So those that have two colors show high likelihood in two methods, and from those, the ones that also fall inside the respective criteria show a higher likelihood in all of the mentioned methods. The shaded area shows the selection in the corresponding plane with the same color-code as mentioned. In the top panel, different limiting sSFR for SED based selection is shown at different redshifts. }
    \label{fig:comparison}
\end{figure*}

To each galaxy, we assigned a likelihood measure that is the median of the likelihoods estimated for that galaxy from each of the three techniques: the UVJ selection (Section \ref{uvj}); Observed color selection (Section \ref{observed-color}) and SED fitting method with and without the nebular emission contribution (Section \ref{sed-fitting}). We combine the results from the SED fitting under different assumptions using the weighted average of the likelihoods for each galaxy using the marginal likelihood calculated for a particular model as their corresponding weights. By doing this, we make sure that we have put more emphasis on likelihoods calculated from the models with higher marginal likelihoods. Then we take the median of all these methods as our final indicator. We select the final sample as those with the median likelihood higher than $0.5$. However, we use the $0.3$ and $0.7$ criteria as our least and most conservative samples, respectively. By using the median indicator, we limit our sample to those galaxies that were assigned at least by two of the methods mentioned to have a high likelihood of being a massive evolved galaxy. Figure \ref{fig:n_density_bounds} shows how this threshold changes the number density measurements of each selection method as well as the median value which is taken as the final indicator. The final selected galaxies with their assigned likelihoods and estimated physical parameters are listed in Table \ref{table:cand}. The galaxies that were selected with the median likelihood higher than 0.3, constitute the most inclusive sample and the one used for finding the upper bounds on the number/mass densities. Also, we control the possible contamination in our final sample, since, we rely on the composite indicator compared to a single measure. Figure \ref{fig:comparison} shows a comparison between different selection methods and how they relate to each other (All the highlighted points are galaxies in our final sample). For example, the selection in the sSFR vs. $M_s$ plane is generally consistent across different methods since the objects with higher BBG/UVJ likelihood are close or inside the selection boundary. For selection in the UVJ colors, we have several candidates that are far from the boundary but show much higher BBG/SED likelihood. This shows the sensitivity of final results to the libraries used, and by changing the SED fitting code and/or the libraries used, we are not necessarily searching through the same part of the models' parameters space. In terms of the selection based on the observed colors, we tend to have higher UVJ/BBG likelihood, but there are a couple of candidates that show higher UVJ/SED likelihoods but fall outside of the criteria. This shows the importance of using information from other bands as well. 

\section{Photometric Uncertainty} \label{Photometric_errors}
All the methods discussed in the previous section are sensitive to uncertainties in photometry at different levels. To quantify the sensitivity of each of the selection methods to photometric uncertainties, we use the Monte Carlo resampling and produce different realizations of each galaxy's photometries. We quantify the confidence level for each candidate galaxy selected from different methods, and the error estimates that propagated into the selections.

Assuming a Gaussian error distribution for each band, we produce realizations for each galaxy by perturbing the photometric measurements within their assigned uncertainties. In other words, we build a bootstrap resampled ensemble of photometric measurement distribution of a galaxy. Then we perform the methods for selecting candidate galaxies and study how these perturbations affect our selections and how sensitive the selected sample is to changes in photometric measurements within their uncertainties. In the case of the selection based on the observed color technique, we only look at several bands, so the uncertainties on the other bands (not used in the selection) are not crucial, but in the case of the UVJ and SED we use all the photometric data which means that the uncertainties in every band can potentially be imperative. 


\subsection{Uncertainty of selection based on UVJ}
Here we consider the effect of the photometric uncertainties in finding UVJ colors and how that could affect the final selected sample. We first choose a subset of the selected galaxies, construct 50 realizations of each, and run SED fitting on the resampled galaxies. Then we quantify how these uncertainties propagate into our inferred rest-frame colors and how they change the result from the UVJ selection in Section \ref{uvj}.

\begin{figure*}
  \vspace{0.5cm}
    \includegraphics[scale=0.41]{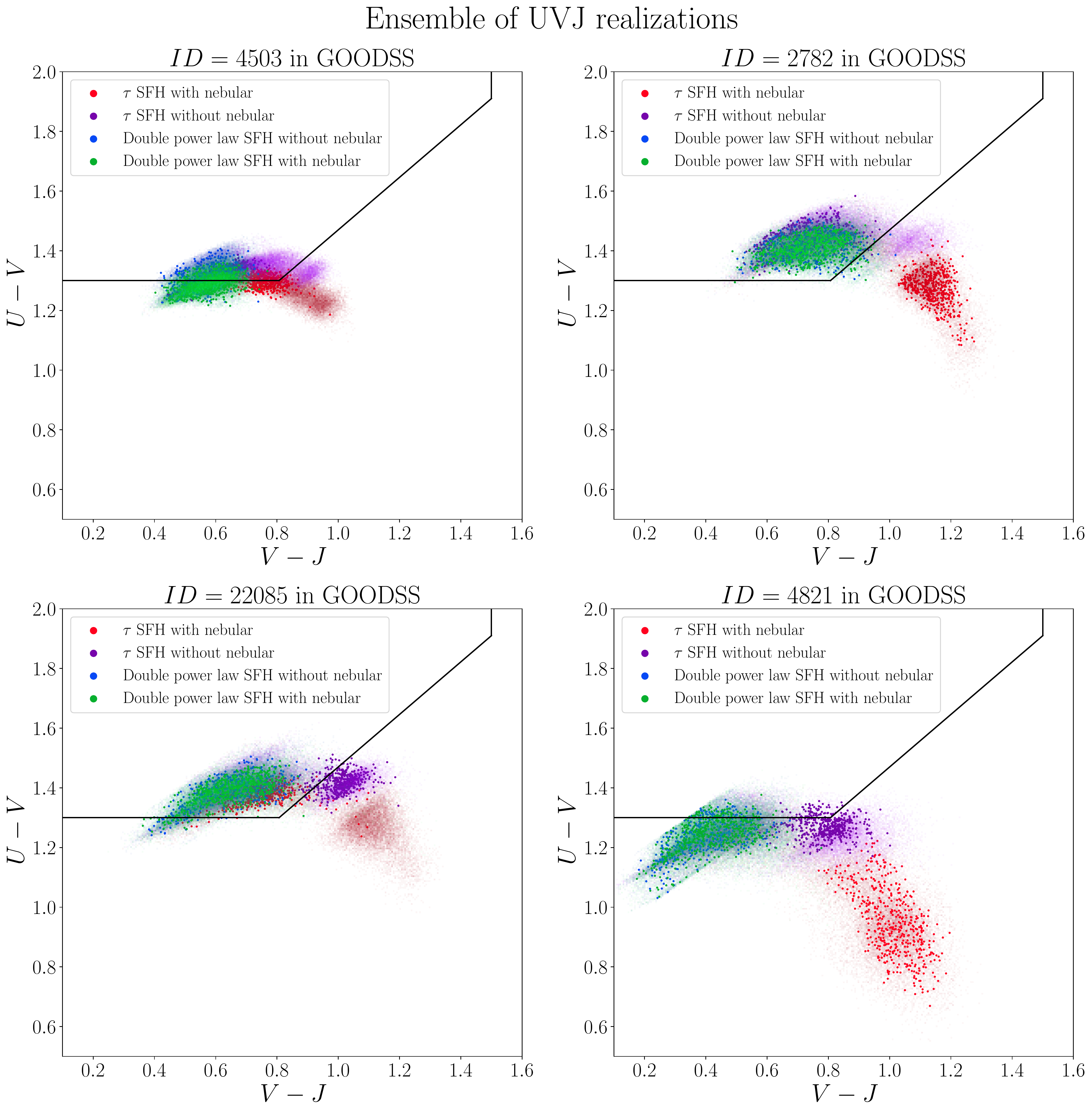}
  \caption{The UVJ plane for galaxies resampled ensemble (Monte Carlo simulated photometry). The UVJ colors from the unperturbed photometry are drawn with larger points, and the dimmer points represent the posterior UVJ color for the disturbed photometries on the UVJ color plane. This is a zoomed-in version of the UVJ plot showing the separation of different models. This figure shows how the change in the model assumptions (i.e., star formation histories, including nebular emission) can change the UVJ colors posterior distribution, which can affect our selection based on the UVJ colors.}
    \label{fig:uvj_realization}
\end{figure*}

Figure \ref{fig:uvj_realization} shows the effect of photometric uncertainty on the UVJ selection. It is clear how the choice of the SFH and whether to include the nebular emissions are pivotal in selecting the quiescent population. We find the UVJ colors from the \textsc{Bagpipes} SED fitting code since we are interested in understanding the effect of the photometric uncertainty on the inference from the SED, which makes a Bayesian posterior estimate more reliable given the clear uncertainty definition based on the posterior. For all but a small subset of galaxies fitted with $\tau$ models and nebular lines, the posterior UVJ of the resampled galaxies falls around the posterior from the true photometry with considerable scatter. For a small subset of galaxies (5 out of 28) fitted with $\tau$ models, the locus of the posterior varies dramatically, and with higher scatter than the posterior UVJ from the true photometry. The double power-law model is found to be more immune to the photometric uncertainties, as is also the case for physical parameters used in SED based selection.

\subsection{Uncertainty of selection based on observed colors} \label{errors-observed}
\begin{figure}[h!]
    \includegraphics[scale=0.54]{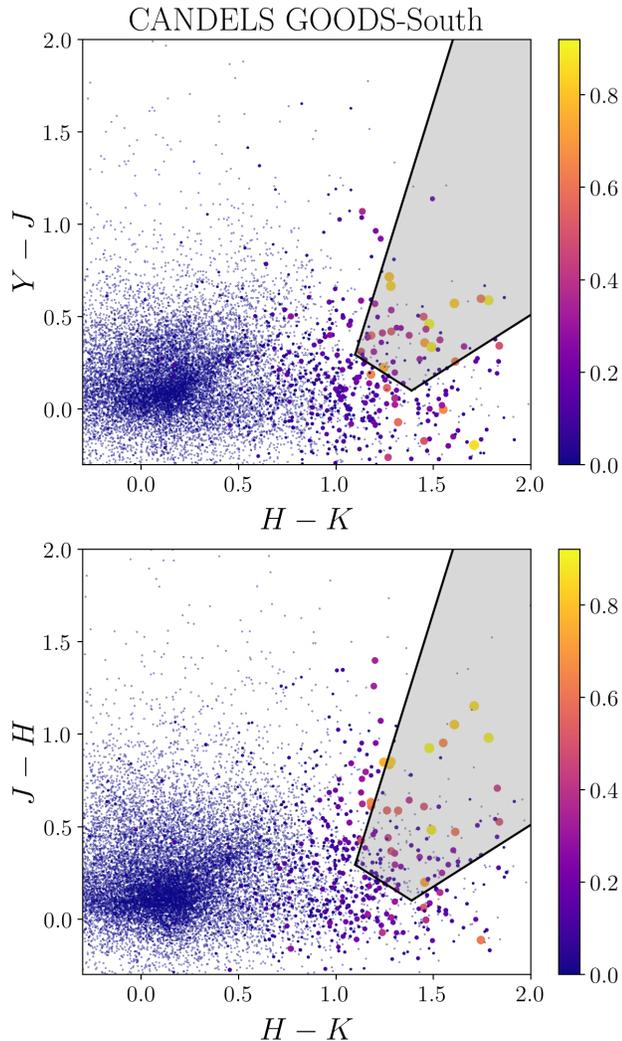}

  \caption{The figure shows $J-H$ vs. $H-K$, and $Y-J$ vs. $H-K$ plane of CANDELS in the GOODS-South field. The larger and more yellow data points show more chances of being quiescent given the photometry measurements and their uncertainties. Here we show just the effect of the photometric uncertainty.}
    \label{fig:bbg_resampled}
\end{figure}

To find the effect of photometric uncertainties on the selection based on the observed colors, we generate $10^5$ realization for every galaxy and assign a likelihood based on the number of realizations that fall into the hard selection criteria to the total number of realizations. The results for the GOODS-S galaxies are shown in Figure \ref{fig:bbg_resampled}, where they are colored and resized based on the assigned likelihood. Few candidates fall outside of the selection criteria, but when we include the photometric uncertainties, we find realizations that fall in the selected region. On the other hand, few galaxies fall inside the criteria, but because of their photometric uncertainties do not show a clear indication of being quiescent.

\subsection{Uncertainty of SED fitting based selections}
With having a posterior distribution of every physical parameter, we can estimate the uncertainties of these parameters. We use the resampling method mentioned above to get even more reliable error estimates. (more details on the effect of the sampling error in \citealt{Higson2018})     

\begin{figure*}
\centering
    \includegraphics[scale=0.48]{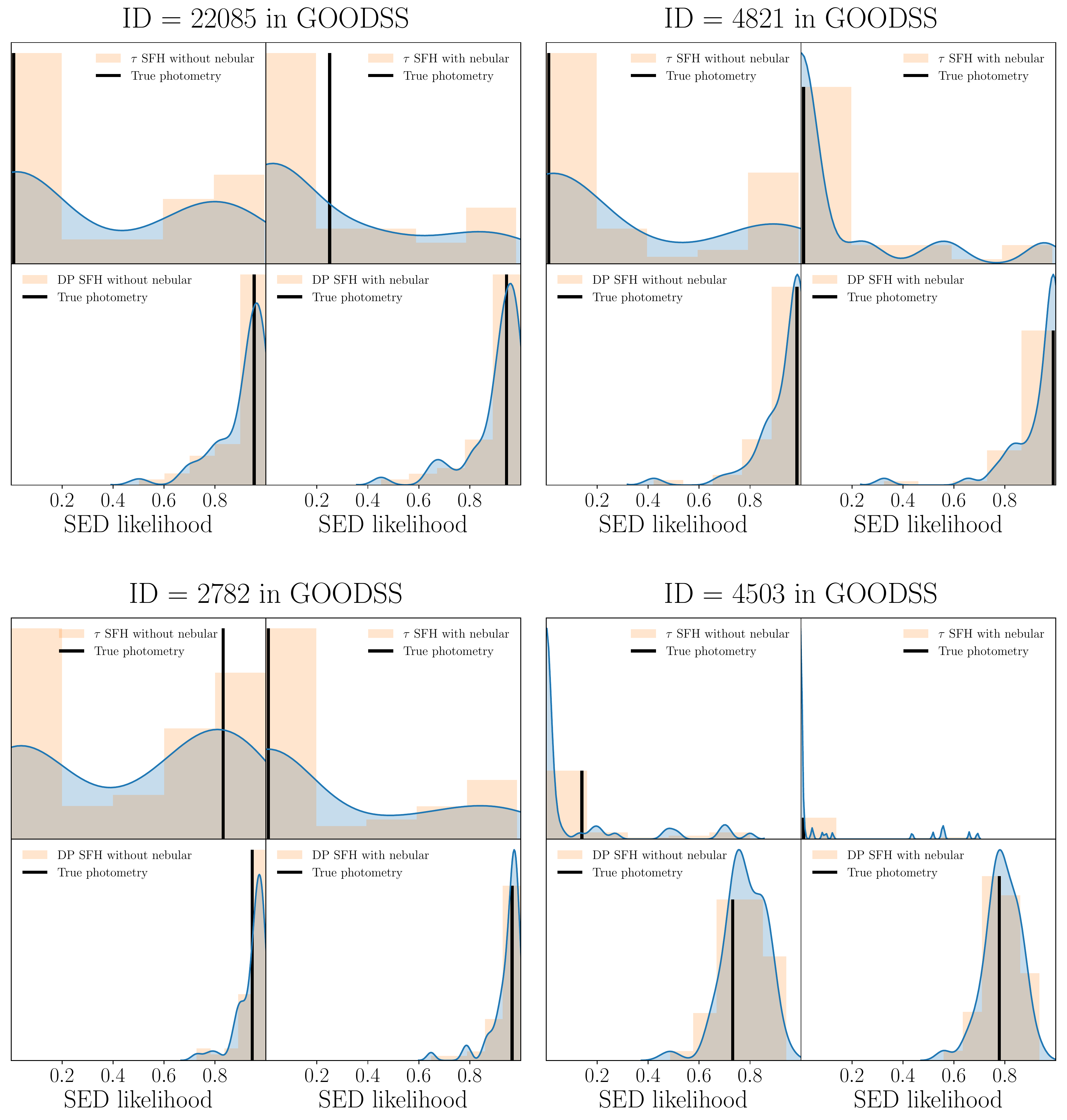}
  \caption{Here we show the dependence of the likelihood assigned in the SED selection on photometric uncertainties. $\psi_{SFR}$ likelihood for several galaxies and their bootstrapped photometries assuming Gaussian uncertainties. For each galaxy, the top left and right is $\tau$ model without and with nebular emission respectively, and the bottom left and right is double power-law without and with nebular emission, respectively. The histograms show bootstrapped sample photometries, and the curve is their simple kernel density estimate. The black line shows the quiescent likelihood from SED from true photometry. The likelihood estimate from $\tau$ models show to be more sensitive to possible variations under photometric uncertainties.}
    \label{fig:sfr}
\end{figure*}

Here we used the same sample from the SED fitting results for finding the sensitivity of the UVJ colors on photometric uncertainties. We follow the effect on the posterior $\psi_{SFR}$, which is the critical parameter in the SED based selection discussed in Section \ref{sed-fitting}. As Figure \ref{fig:sfr} shows, for most of the sample used, the $\tau$-model SFH is more susceptible to photometric uncertainties than double power-law, which has more stable results under photometric perturbations similar to the UVJ colors. Also, including the models' nebular emission leads to models with UVJ colors that are further away from the quiescent region and hence reduce the number densities of the quiescent population.

In this section, we found the effect of the photometric uncertainties on each method's assigned likelihood. As mentioned in Section \ref{observed-color}, for finding the likelihood based on the observed colors for all the galaxies in the CANDELS fields, we assign a likelihood to every photometric realization of a galaxy based on their proximity to the selection boundary, then we take the average of these as the observed color likelihood for that galaxy. By doing this, we can take into account both the photometric uncertainties and closeness to the selection boundary. Since we only calculated the effect of the photometric uncertainties on likelihoods based on the UVJ and SED based selection for a sub-sample of galaxies, we did not incorporate the photometric uncertainties in the final assigned likelihood of these methods.

\section{Number and Stellar Mass Densities}\label{number-densities}
\begin{figure}[!ht]
    \includegraphics[width=0.47\textwidth]{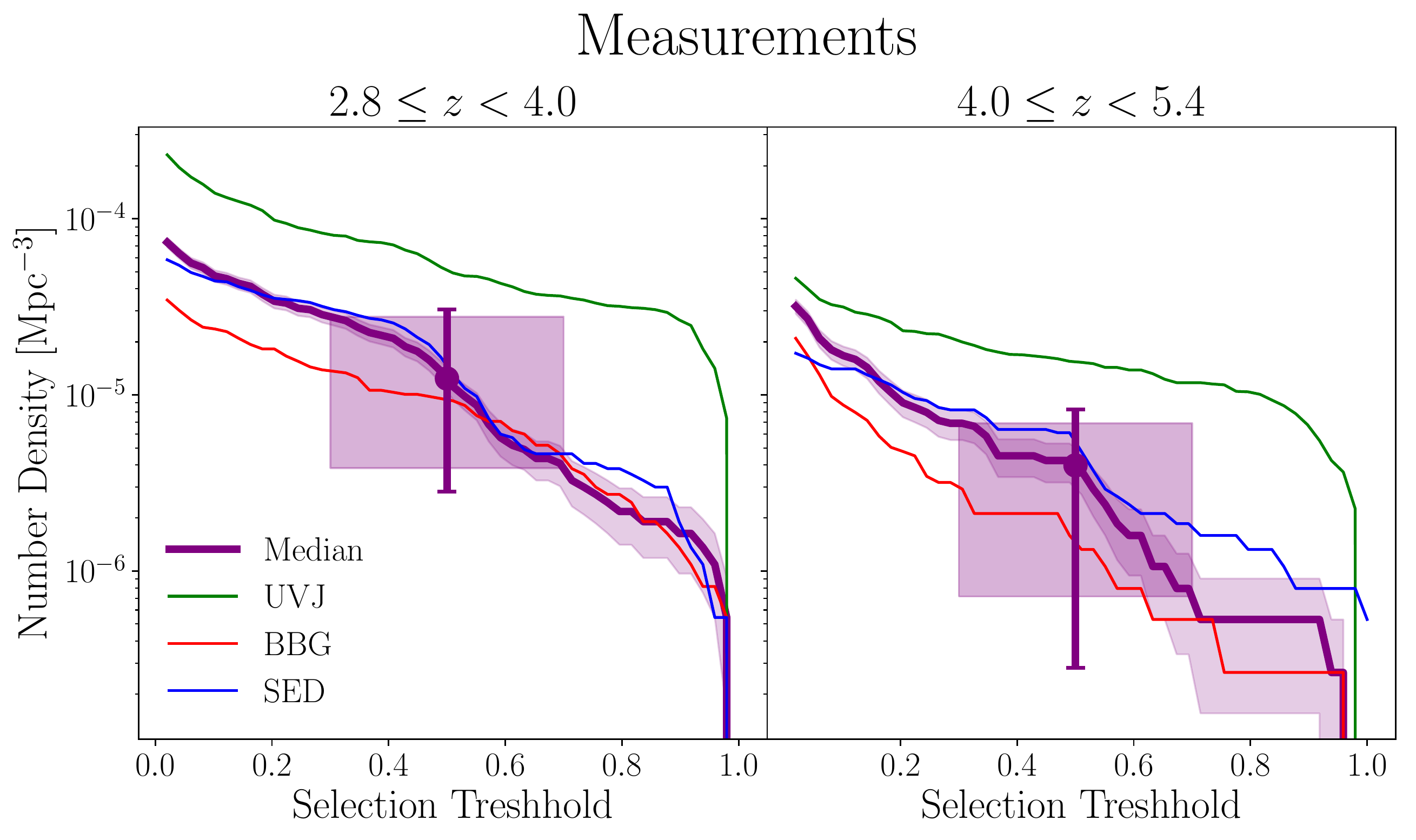}
    \caption{The comoving number densities for different selection methods and the median indicator used for the selection of the final sample are plotted. The shaded area around the median line shows the 1 $\sigma$ Poisson noise at a given threshold. We define the upper bound as the number density at threshold 0.3 and the lower bound as 0.7, including the Poisson noise. The error bars show the uncertainty as upper and lower bounds. The shaded rectangular correspond to variations in the number densities due to variation in the selection threshold ($[0.3, 0.7]$).}
    \label{fig:n_density_bounds}
\end{figure}

Using the final sample of massive and evolved galaxies, as presented in Table \ref{table:cand}, we now estimate their number and mass density across the CANDELS fields. We divide the candidates based on their redshifts into two bins $2.8 \leq z < 4.0$ and $4.0 \leq z < 5.4$, approximately representing the same comoving volumes. To find the most reliable candidates, we need to assign a threshold for our likelihood. However, in order to follow this threshold's effect, we impose three different values on the final likelihoods. These thresholds are shown as error bars in Figure \ref{fig:n_density}. The upper limits are for galaxies with an assigned likelihood of more than 0.3, the lower limits are for those with likelihood more than 0.7, and the measurement points are for those more than 0.5. The variation caused by changing the selection threshold is larger than the typical Poisson noise. However, we take into account the Poisson noise as an independent source of uncertainty (Figure \ref{fig:n_density_bounds}). We find the number/stellar mass densities after taking into account the incompleteness of our sample, as discussed in the next section.

\subsection{Completeness}
To estimate the completeness of the sample used, we follow the prescription adopted in \citealt{Pozzetti2010}. We divide the CANDELS galaxies into two redshift bins mentioned above. We find the 20\% of the galaxies with the faintest apparent magnitude in H Band, with specific star formation rates lower than 75\% of galaxies in that redshift bin. We calculate the $M_{lim}$ defined as the mass which the galaxy would have if it were at the survey limiting magnitude ($H_{lim} = 26$) (assuming a constant mass-to-light ratios); in other words, we find $\log (M_{lim}/M_*) = 0.4 (H - H_{lim})$. Then we define the fraction of the sample with $\log(M_{lim}/M_\odot) > 10$ as the completeness of the quiescent galaxies with that mass range. The effective volume of the survey at each redshift bin is defined to be $V_{eff} = V_{co} \eta(z)$; where $V_{co}$ is the comoving volume at each redshift bin and $\eta(z)$  is the completeness of the sample for $\log(M_{*}/M_\odot) > 10$ at each redshift bin of the quiescent population.

\subsection{Comparison to models predictions}
In this section, we find the target population predicted from the several existing models listed below and compare their number and stellar mass densities with the sample found in this study.

\begin{itemize}
    \item {\textbf{Simulated Infrared Dusty Extragalactic Sky:}
We use the simulated catalog presented in \citealt{Bethermin2017}, which uses the abundance matching technique for occupying dark matter halos from the Bolshoi-Planck simulation (\cite{Klypin2016}; \cite{Rodr2016}) and uses an updated version of the 2SFM (two star-formation modes) galaxy evolution model (\cite{Sargent2012}; \cite{Bethermin2012} from which a lightcone covering 2 degrees squared was produced.}

    \item {\textbf{Universe Machine:}
We employ the available CANDELS lightcones produced by the Universe Machine semi-empirical model, which uses the same Bolshoi-Planck simulation for halo properties and mass assembly histories. There are eight realizations for each field in CANDELS, and we use all of these realizations. (\citealt{Behroozi2019UM})}

    \item {\textbf{\textsc{Eagle} Simulation:}
We use the \textsc{Eagle} cosmological hydro-dynamical simulations of a box with $L = 50$ and $100$ co-moving mega-parsec (cMpc) on each side. We use \textsc{RefL0050N0752}, \textsc{RefL0100N1504}, and \textsc{AGNdT9L0050N0752} in which the first two are the reference physical model used and the last one has higher AGN heating temperature and lower sub-grid black hole accretion disc viscosity. All of these simulations have the same mass resolutions. (\citealt{Schaye2015}; \citealt{Crain2015}; \citealt{McAlpine2016}; \citealt{EAGLE2017}) }

    \item {\textbf{\textsc{IllustrisTNG} Simulation:}
We use the latest \textsc{IllustrisTNG} cosmological hydro-dynamical simulations of a box with $L = 100$ and $300$ co-moving mega-parsec (cMpc) on each side. There are three resolutions for each simulation box, and here we used the highest resolution. (\citealt{Nelson2018}; \citealt{Springel2018}; \citealt{Pillepich2018}; \citealt{Marinacci2018}; \cite{Naiman2018}; \citealt{Nelson2019})}
\end{itemize}

To identify the quiescent galaxies in the simulations and to be consistent with observation of the massive evolved galaxies at $z>2$ which has been known to be quite compact $R_e \leq 2.5$ kpc (\citealt{Daddi2005}; \citealt{Trujillo2006}; \citealt{Trujillo2007}; \citealt{vanDokkum2008}; \citealt{Newman2010} \citealt{Damjanov2011}; \citealt{vanderWel2014}), we use the stellar mass and star formation rates within the twice of half-mass radius. \citealt{Donnari2019} has shown that the star-formation main sequence in the \textsc{Illustris-TNG} 100 for $z>2$ is quite identical when assuming physical properties within an aperture of $5$kpc compare to the twice the half-mass radius definition for both $10, 1000$ Myr timescale of SFR measurements. Also, following \cite{Merlin2019} and staying consistent in our comparison within simulations, and since the physical properties of the \textsc{Eagle} simulations are available within certain apertures in the range $1$ to $100$ kpc, we use the physical properties within an aperture that is closest to four times the half-mass radius. For all the models we select galaxies with stellar masses $\log(M_{*}/M_\odot) > 10$, and specific star-formation rate within the half-mass radius lower than the evolving sSFR constraint employed in the \cite{Pacifici2016}, as $sSFR_{lim} = 0.2/t_U(z)$, where $t_U(z)$ in Gyrs is the age of the universe at redshift $z$. \cite{Carnall2019b} showed this to be consistent with the $\psi_{SFR}$ measure used in this work across different redshifts.

\begin{table*}
\centering
\begin{tabular}{ccccccccc}
\hline
\textbf{Redshift bin} & \textbf{Completeness} & \textbf{Number Density} ($Mpc^{-3}$) & \textbf{Upper bound} ($Mpc^{-3}$) & \textbf{Lower bound} ($Mpc^{-3}$)\\

\hline
$2.8\leq z < 4.0$  &  0.8 & $1.2 \times 10^{-5}$ & $3.1 \times 10^{-5}$ & $2.8 \times 10^{-6}$\\

$4.0\leq z < 5.4$ &  0.5 & $4.1 \times 10^{-6}$  & $8.2 \times 10^{-6}$ & $2.7 \times 10^{-7}$\\

$2.8 \leq z< 5.4$ & 0.7 & $8.1 \times 10^{-6}$ & $1.9 \times 10^{-5}$ & $1.7 \times 10^{-6}$ \\
\hline
\end{tabular}
\caption{The number density measurements and calculated upper and lower bounds (before completeness correction) used in the Figure \ref{fig:n_density} and \ref{fig:n_density_bounds}. Lower bound and upper bound correspond to the selection threshold of 0.7 and 0.3 respectively. Poisson errors are included in the bounds.}
\label{table:number_density}
\end{table*}

\begin{figure}[!ht]
    \includegraphics[width=0.46\textwidth]{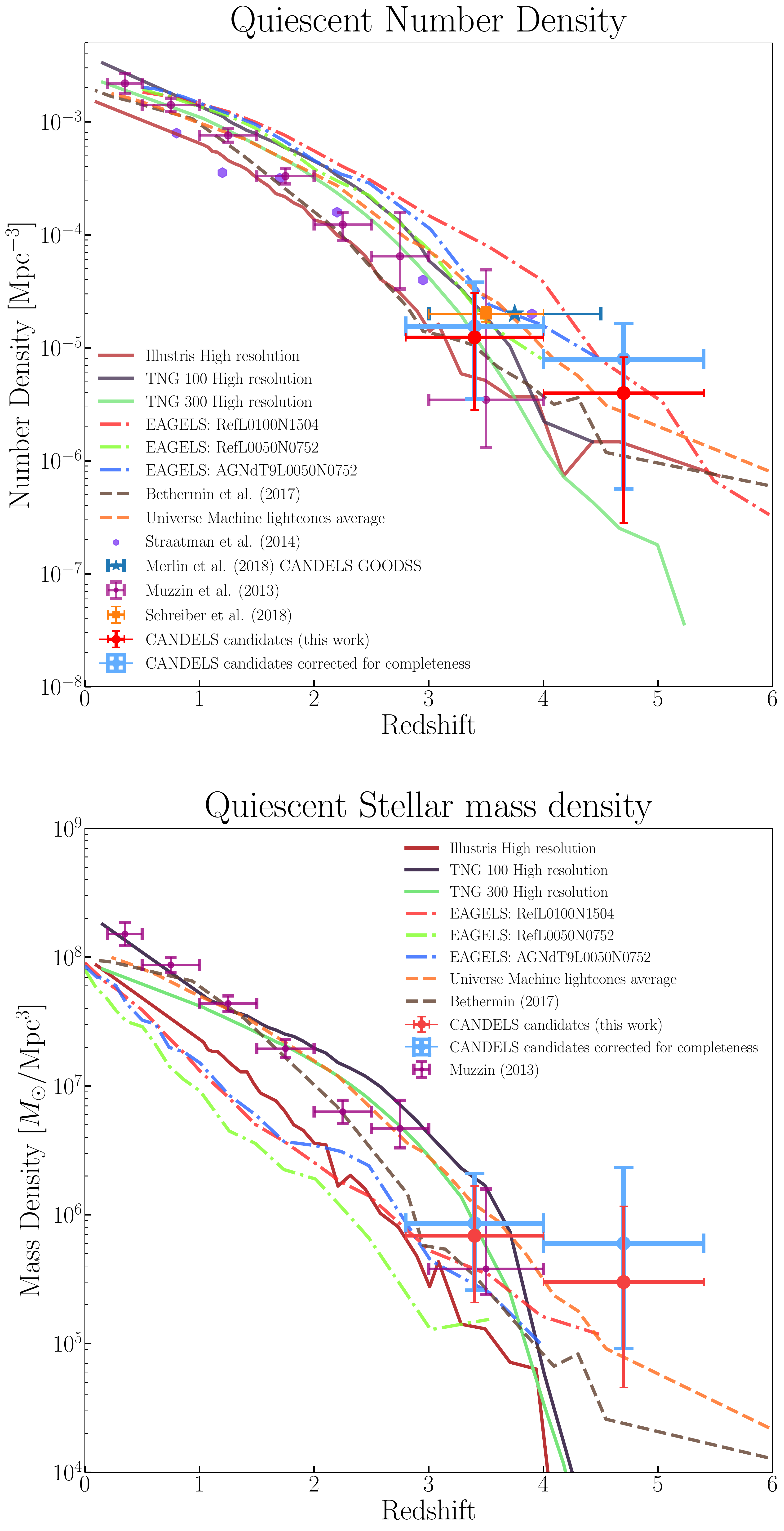}
    \caption{The comoving number and stellar mass density of the massive quiescent galaxies (defined as those with $M_s \geq 10^{10} M_\odot$ and $sSFR \leq 0.2/t_H(z)$ Gyr$^{-1}$ in which the $t_H$ is the age of the universe at the redshift of $z$) in two redshift bins from this work are overplotted as red data points on those from the Illustris TNG, and EAGLE simulations for different volumes are shown. The average of the Universe Machine (\cite{Behroozi2019UM}) for all the CANDELS's lightcone realizations is the orange dashed line. The model from \cite{Bethermin2017} is shown with the deep brown dashed line. Measurements from \cite{Muzzin2013}, \cite{Straatman2014}, \cite{Schreiber2018}, and \cite{Merlin2018} are shown as purple, violet, orange, and blue data points, respectively. The error bars show one $\sigma$ uncertainty in measurements, except for the error bars of this work, which are upper and lower limits of the measurements, taking into account the dependence of the measurement on artificial selection thresholds and the one $\sigma$ Poisson uncertainty.}
    \label{fig:n_density}
\end{figure}

\begin{figure}[!ht]
\vtop{%
  \vskip-1ex
  \hbox{%
    \includegraphics[scale=0.26]{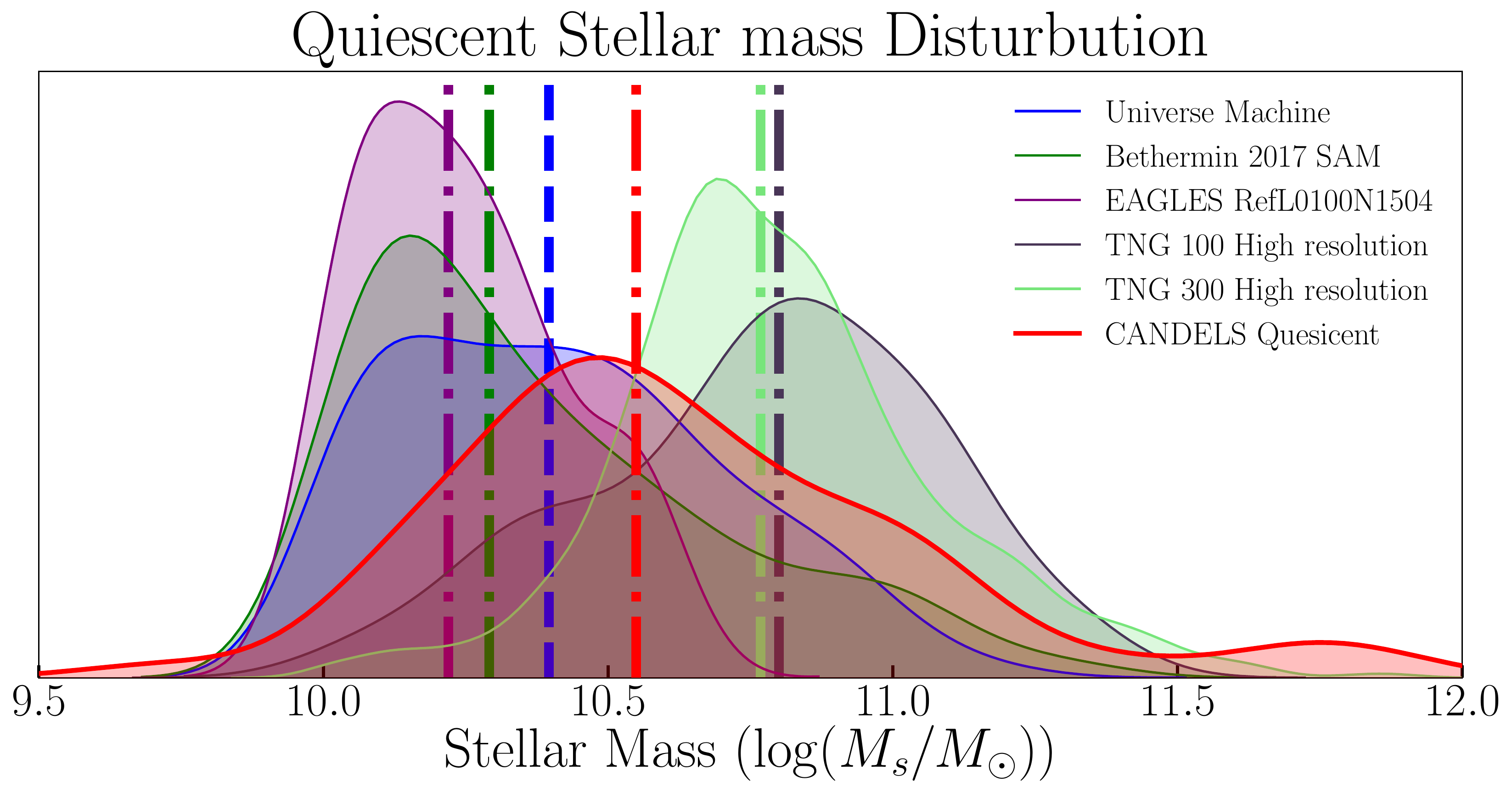}  }%
}

    \caption{Shows the Stellar Mass distribution of the candidates found here compared with semi-analytical models and hydro-dynamical simulations. The y-axis shows the kernel density estimate of the relative frequency (Using Gaussian kernel and Scott's rule for bandwidth selection). The dashed line shows the median of each sample.}
    \label{fig:hist_mass}
\end{figure}

Figure \ref{fig:n_density} shows the comparison between the observed number density of the quiescent galaxies within the corresponding comoving volumes, corrected for their completeness. The observed and predicted counts from the numerical simulations and empirical models of galaxy evolution are in generally good agreement. However, at the higher-end of the redshift distribution, there are more observed candidates than predicted by models by a factor of $\sim 5$. Nevertheless, it is somewhat consistent with the uncertainties shown in Figure \ref{fig:n_density} (especially in the case of Universe Machine). Additionally, the presence of possible contaminants in the sample can ameliorate the tension.

\begin{figure*}[!ht]
\vtop{%
  \vskip-1ex
  \hbox{%
    \includegraphics[scale=0.4]{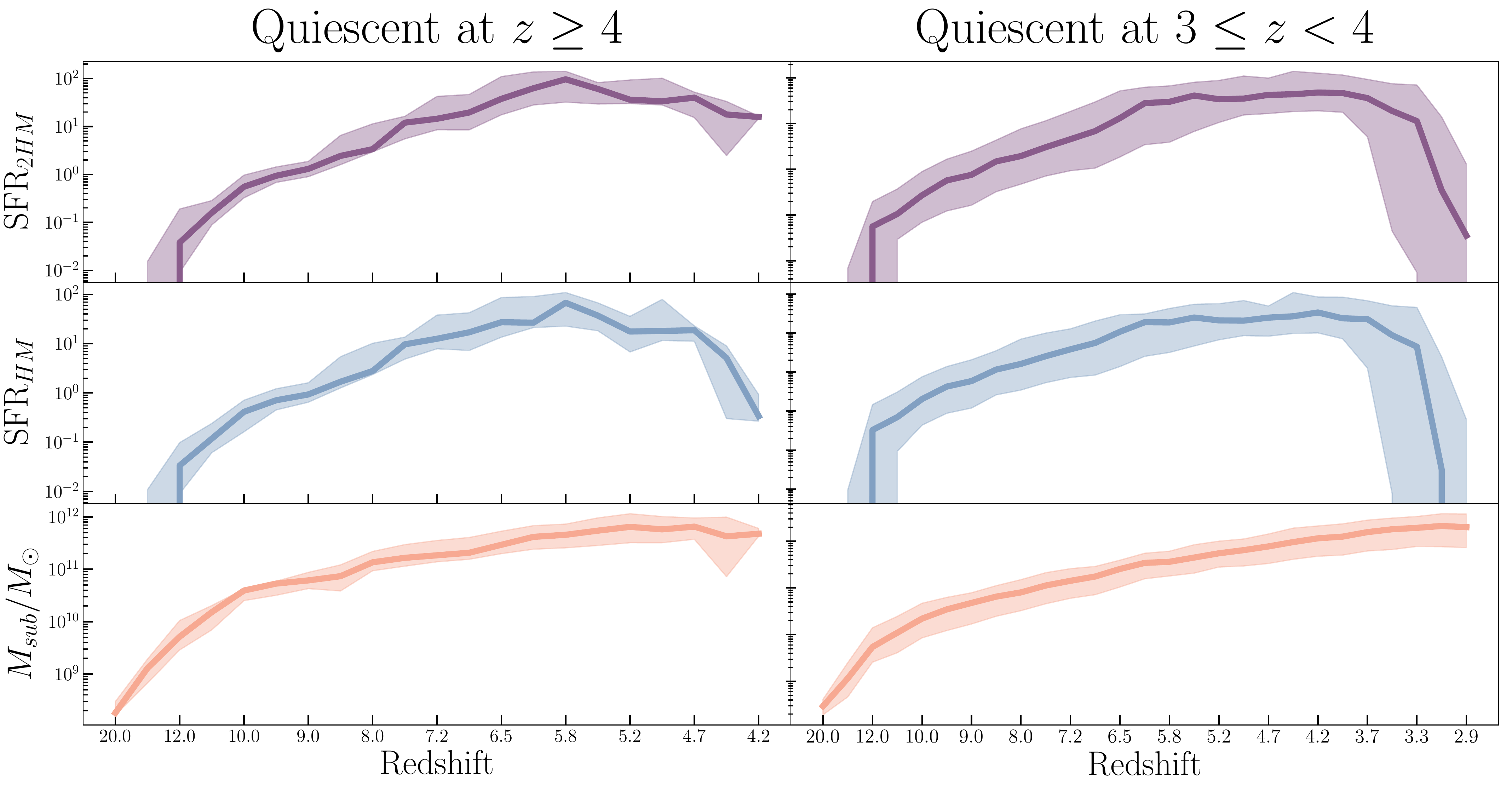}  }%
}

    \caption{Shows the history of total subhalo mass assembly, star formation rate within the half-mass radius (SFR$_{HM}$), and within twice the half-mass radius (SFR$_{2HM}$) from the \textsc{IllustrisTNG}100-1 simulation merger trees and following the most massive progenitor branch for the selected sample of quiescent galaxies based on the stellar mass and SFR measurements defined within the half-mass radius.  The solid lines show the median history with the shaded area corresponding to the 20\% and 80\% percentiles. As the figure shows, the higher redshift bin candidates show that the SFR within the half-mass radius is decreasing more rapidly than the star-formation within twice the half-mass radius as they evolve.}
    \label{fig:SFH_TNG}
\end{figure*}

Figure \ref{fig:n_density} further shows the stellar mass density for our sample compared to the simulations and empirical models. We used the mass catalog from CANDELS to measure the stellar mass density. The tension between the stellar mass density for our sample with the simulations seems to be worse; however, the argument above for the number density applies here as well. The stellar mass measurements from the SED fitting results when including nebular line emission can also mitigate the tension as the presence of the nebular lines can reduce the contribution from the continuum part of the SED, which can result in lower inferred stellar masses. Also, by looking at the mass distribution of the selected samples from models and our CANDELS candidates (Figure \ref{fig:hist_mass}) shows that the CANDELS candidates are relatively more massive than their counterparts in models except for those from the TNG simulations. The TNG100-1 sample shows to be more massive and using the fact that the stellar mass density at the highest redshift bins is lower than what we found here, we argue that the sample galaxies from the TNG tend to build their stellar mass later and at a higher rate than what is suggested from the CANDELS sample similar to what \cite{Fontana2009} found. Also, by looking at the merger tree information of the selected sample (\citealt{Rodriguez2015}) and following their history back to their formation epoch, we find the mass assembly and star formation histories of the selected sample to check whether it is consistent with their selection as quiescent galaxies. Figure \ref{fig:SFH_TNG} shows the median and (20\%, 80\%) percentile of the mass assembly and star formation rate history of the sample selected. Star formation histories were calculated from two SFR measures, one within the half-mass radius and one within twice the half-mass radius. Figure \ref{fig:SFH_TNG} reveals that the TNG sample at the lowest redshift bin is consistent with quenching even within the twice of the half-mass radius. Although the higher redshift sample shows modest evidence of quenching within the half-mass radius, the star formation seems to continue outside of the half-mass radius.


\section{Discussion and Conclusion}\label{discussion}
In this study, we used three methods of selecting quiescent galaxies at high redshift. Using the grading scheme explained in Section \ref{selection-methods}, we minimized the effect of selection criteria by assigning a likelihood measure to each galaxy under each selection method. Figure \ref{fig:n_density_bounds} shows the number density measurements from different methods and the composite indicator (median) we used in the final selection and their dependence on the selection thresholds. We also investigated the effect of the photometric uncertainties to understand the degree to which the result represented here could be affected. We combined these likelihoods by taking the median across different methods. The results from this study are consistent with similar investigations in CANDELS fields. \cite{Nayyeri2014} found 16 candidates in the GOODS-South of which seven are the same as what we found as candidates. Also, \cite{Merlin2019} found 102 candidates across all the CANDELS, with 36 being the same as those found here. More than half of the proposed candidates that are missing from our sample but reported in \cite{Merlin2019} showed more substantial Bayesian evidence when they were fitted with $\tau$ SFH, and double power-law SFH compared to the constant star-formation histories employed in \cite{Merlin2019}. This indicates the selection is sensitive to the form of the SFH assumed.

We do not expect a significant difference between the two methods since UVJ colors are expected to follow the sSFR of the model galaxies. However, they react differently to somewhat arbitrary criteria either on the UVJ plane or on the $sSFR$. Moreover, we showed the classification based on the UVJ colors and SEDs to be strongly dependent on the choice of the star formation history and, in the case of $\tau$ model library, are particularly susceptible to photometric uncertainties.

We found no strong evidence in favor of any star-formation history, similar to what found in \cite{Carnall2019b}. However, double power-law showed more substantial evidence, with less sensitivity to photometric uncertainties. Also, models that included nebular emissions showed stronger evidence. However, in the grading scheme defined here, we weighted each galaxy's likelihood by its respective evidence so that different models contribute to the likelihoods based on their relative Bayesian evidence. 

The selection based on observed colors only uses a few passbands and does not require a search through the parameter space to find the best-fitted model. Therefore, it is computationally more feasible than SED or UVJ based selections when a measure of the sensitivity to photometric uncertainties is needed. \cite{Hemmati2019} has proposed a faster method of measuring the physical properties of galaxies and their uncertainties using Kohonen's Self-Organizing Map (SOM) (\citealt{Kohonen1982}). SOM is a neural network that maps the entire color space of galaxies into lower dimensions while preserving the topology of the input space (\citealt{Kiviluoto1996}; \citealt{VILLMANN1999}). This feature ensures that we can have measurements with stable behavior under photometric uncertainties. 

We find that the number and stellar mass densities of the candidates are generally consistent with the prediction of the numerical simulations. However, the agreement becomes less pronounced at higher redshifts, where numerical simulations underpredict the number/stellar mass densities of the quiescent galaxies.

Although final sample purity requires spectroscopic observations, we employ a conservative approach for the completeness correction used in our number and stellar mass density measurements. In the sense that, if we had used the model galaxies with a quiescent stellar population with \textsc{Fsps} (\citealt{Conroy2009}; \citealt{Conroy2010}) \footnote{Using $\tau$ model star-formation histories with $\tau=0.2$ Gyr and the age equal to the age of the universe at each redshift bin, with solar metalicities.}, we get completeness levels that would make the number and mass densities even higher than what we showed here, especially at higher redshift bin. 

Using the number densities found here and an abundance matching technique, we predict an upper limit for the lowest halo masses of these galaxies and the corresponding lower limit for the highest efficiency of galaxy formation (\citealt{Vale2004}; \citealt{Wiklind2008}; \citealt{Conroy2010}; \citealt{Guo2010}; \citealt{Moster2010}; \citealt{Behroozi2010}; \citealt{Trujillo-Gomez2011}; \citealt{Reddick2013}; \citealt{Moster2013}). Here, we use the abundance matching technique described in \cite{Behroozi2013b}, which matches galaxies in the order of decreasing stellar mass to the decreasing peak historical halo mass, according to their accretion history. Therefore, we find the peak historical halo mass corresponding to the number density of candidates, using cumulative peak historical mass function from \cite{Behroozi2013a} at a given redshift. We find the corresponding dark matter halos of mass $M_h \approx 4.2, 1.9, 1.3  \times 10^{12} M_\odot$ at different redshift bins in Table \ref{table:number_density} centered at $z=3.4, 4.1, 4.7$ and $M_h \approx 6.5, 4.7, 4.5 \times 10^{12} M_\odot$ for the lower bound of the number densities, respectively. The estimated halo masses are similar to the critical shock-heating halo mass at these redshifts (\citealt{Birnboim2003}; \citealt{Kere2005}; \citealt{Cattaneo2006}), at which the transition from cold-mode accretion into partly hot-mode accretion happens \cite{Dekel2009}. This transition could explain part of the quenching mechanism simply by shutting off the accretion of most of the cold gas into the halo. Although the fraction of the cold/hot mode accretion is higher than in the halos of the same size at a lower redshift, the halo masses found here suggest a relatively lower fraction of cold/hot mode accretion than the less massive halos at the same epoch. 

This agrees with the ``cosmological starvation'' scenario as a quenching mechanism, that suggests there is a tight correlation between the star-formation rates and the mass accretion rate onto the halo (e.g., \citealt{Larson1980}; \citealt{Balogh2000}; \citealt{VanDenBosch2008}) and was shown to be able to explain the relatively quick quenching time for massive central galaxies at high redshift (Argo simulations, \citealt{Feldmann2015}). The relatively high star formation rate required for building the stellar mass within a short timescale is consistent with sub-millimeter galaxies, that exhibit high star-formation rate of order of few $100$ $M_\odot$ (e.g., \citealt{Marchesini2010}; \citealt{Swinbank2013}), as a potential progenitor of these post-starburst systems (e.g., \citealt{Toft2014}; \citealt{Wild2016}; \citealt{Wild2020}).

Therefore, we suspect that a part of the quenching scenario of these systems could be that there was a phase of high accretion of gas into halos, which leads to high levels of star-formation rates and the subsequent mass build-up. Then the accretion of cold gas declines as the starvation phase begins, possibly by transiting to hot-mode accretion of the gas into the halo, followed by an ``over-consumption'' phase (\citealt{McGee2014}; \citealt{Balogh2016}), in which massive galaxies at high redshift use their gas in a depletion time scale which leads to their relatively fast quenching (in $\sim$ few $100$ Myrs). Similarly, \citealt{Estrada2019E} found that massive galaxies at $z \sim 1-2$ enriched quite rapidly to approximately solar metallicities as early as $z \sim 3$. However, radio mode AGN feedback may still be required for keeping the galaxy quiescent and suppressing any residual star formation by driving out the remaining gas as well as possible later accretion of a cold gas stream that penetrated the hot halo, in combination with stellar feedback and gravitational heating (\citealt{Best2005}; \citealt{Croton2006}; \citealt{Bower2006}; \citealt{Feldmann2015}; \citealt{Man2019}; \citealt{Falkendal2019}). 

Similar to what has been observed at lower redshift, we find the quiescent sample found here to be consistent with the green valley (quenching) transition stellar mass ($log(M_s/M_\odot) \sim 10.7$) going all the way out to $z \sim 5$. This points to the importance of this stellar mass scale in quenching of the galaxies even at higher redshifts. We find that the number density measurements and their corresponding halo masses are consistent with halo quenching as the main mechanism (e.g., \citealt{Rees1977}; \citealt{Blumenthal1984}; \citealt{Kere2005}; \citealt{Dekel2009}). However, certain versions of the AGN quenching models can achieve the same number densities for quiescent galaxies. (e.g., \citealt{Chen2019})

Results suggesting that cosmological starvation plays a crucial role in quenching of the massive quiescent galaxies at high redshifts are based on the abundance matching technique. However, this is based on the halo mass functions, which could underestimate the number of massive halos at high redshifts. The TNG simulations sample of quiescent galaxies and their progenitor history from their corresponding merger tree shows that the primary quenching mechanism is the quasar mode energy injection by AGNs. These point to the AGN activity as the primary quenching mechanism of the candidate galaxies. However, similar to what we found here, it is proposed that the ``over-consumption'' scenario is the primary driver of the environmental quenching for the massive quiescent galaxies at $z \leq 3.5$ (\citealt{Chartab2020}), with massive quiescent galaxies residing in over-dense regions (more than $20$\% more quiescent in over-dense regions), which shows that environment has a crucial role in the cosmological starvation scenario.


\section*{Acknowledgement}
We acknowledge the Virgo Consortium for making their simulation data available. The eagle simulations were performed using the DiRAC-2 facility at Durham, managed by the ICC, and the PRACE facility Curie based in France at TGCC, CEA, Bruy\`eresle-Ch\^atel.


\newpage
\appendix
\section{Selection function}\label{appendix}
To quantify the effect of decision boundaries, we use an extended logistic function to differentiate between points closer to boundaries and those reside well beyond the boundaries. A transformation which takes the distance from the selection boundary and produces a value between $[0, 1]$ in which the value closer to zero represent a galaxy which resides in the star-forming section and far from the selection boundaries and the values close to one is the counterpart inside the quiescent region. However, any squashing function that takes a value from $(-\infty, \infty)$ and maps it to $[0, 1]$ with the limiting behavior discussed above can do the job. We use a generalized logistic function defined below:

\begin{equation}
    Y(x) = A + \frac{K-A}{(C+Q e^{-B(x-M)})}
\end{equation}

\noindent in which $A, K, B, C, Q, M$ are free parameters. However, we can fix several of them without losing the needed features, and by the fact that we are interested in maps from $(-\infty, \infty)$ to $[0, 1]$.
Changing the parameters of this function can change the values we use to distinguish between different candidates. However, the utility of this method is to compare different candidates given a set of chosen selection parameters, and we do not care about a particular value. We are interested in whether a given candidate is more likely to be quiescent than other candidates. A ``reasonable'' choice of these parameters only generalizes the selection methods without losing information compared to applying a  hard selection function. 

To combine different criteria consistently, we assume that the function $Y(x_i)$ is the membership likelihood of the $i^{th}$ galaxy in the fuzzy set defined for specific criteria. ($Y$ is a form of membership function ($\mu$))

\begin{equation}
    A = \{(u, m) \mid u \in U \quad \& \quad m = \mu_A(u)\}
\end{equation}

\noindent in which U is the set of all galaxies, and $\mu_A$ is the membership likelihood of the galaxies to criterion $A$. Also for combining different criterion, we use the following rules for t-norm (intersection) and t-conorm (Union): 

\begin{gather*}
    A \cap B = \{(u, m) \mid u \in U \quad \& \quad m = \mu_{A \cap B}(u)\} \\ 
    A \cup B = \{(u, m) \mid u \in U \quad \& \quad m = \mu_{A \cup B}(u)\} \\
    \\
    \mu_{A \cap B}(x) = \mu_A(x) \mu_B(x) \\
    \mu_{A \cup B}(x) = \mu_A(x) + \mu_B(x) - \mu_A(x) \mu_B(x)
\end{gather*}

\begin{table*}[b!]
\centering
\begin{tabular}{|c|c|c|c|c|c|}
\hline
\textbf{Selection}  & \textbf{Criteria} & \textbf{Source} & \textbf{Set} & \textbf{Likelihood function} &  \textbf{Final criteria}\\

\hline

& $x_1 = 3.44 -(Y-J)-3.40 \times (H-K)$ & &$A_1$ & $\mu_1(x_1)$ & \\
& $x_2 =  - 1.03 + (Y-J) + 0.67 \times (H-K)$ & & $A_2$ & $\mu_1(x_2)$& \\

 & $x_3 = 0.83 + (Y-J) - 0.67 \times (H-K)$ & & $A_3$ & $\mu_1(x_3)$& $((A_1 \cap A_2 \cap A_3)$ \\

Observed colors & $x_4 = 2.48  +  (J-H)- 2.80 \times (H-K)$ & \cite{Nayyeri2014} & $A_4$ & $\mu_1(x_4)$ & $\cup$\\
& $x_5 = - 1.70 + (J-H) +  (H-K)$ & &  $A_5$ & $\mu_1(x_5)$& $(A_4 \cap A_5 \cap A_6))$ \\
& $x_6 = 0.16 + (J-H) - 0.43 \times (H-K)$ & & $A_6$ & $\mu_1(x_6)$& $\cap$ \\
& $x_7 = S/N(U) - 2 $ && $A_7$ & $\mu_2(x_7)$ &  $(A_7 \cap A_8)$\\
& $x_8 = S/N(B) - 2$ && $A_8$ & $\mu_2(x_8)$ & \\
\hline

& $y_1 = (U-V)-0.88 \times (V-J) - 0.59$ && $B_1$ & $\mu_1(y_1)$ & \\
& $y_2 =  (U-V) - 1.2$ & \cite{Whitaker2011} & $B_2$ & $\mu_1(y_2)$ &  \\
 & $y_3 = 1.4 - (V-J)$  && $B_3$ & $\mu_1(y_3)$&  $((B_1 \cap B_2 \cap B_3)$\\
 & $y_4 =   S/N(J) - 2 $ && $B_4$ & $\mu_2(y_4)$ &  $\cup$\\
\cline{2-5}

UVJ & $y_1 = (U-V)-0.88 \times (V-J) - 0.59$ && $C_1$ & $\mu_1(y_1)$ & $(C_1 \cap C_2 \cap C_3)$\\
  & $y_2 =  (U-V) - 1.3$ &\cite{Muzzin2013} & $C_2$ & $\mu_1(y_2)$ &  $\cup$\\
 & $y_3 = 1.5 - (V-J)$  && $C_3$ & $\mu_1(y_3)$&  $(D_1 \cap D_2 \cap D_3))$\\
\cline{2-4}

& $y_1 = (U-V)-0.88 \times (V-J) - 0.56$ && $D_1$ & $\mu_1(y_1)$ & $\cap$\\
& $y_2 =  (U-V) - 1.3$ &\cite{Straatman2014}& $D_2$ & $\mu_1(y_2)$ & $B_4$\\
& $y_3 = 1.6 - (V-J)$  & &$D_3$ & $\mu_1(y_3)$& \\
\hline
\end{tabular}

\begin{tabular}{|c|c|}
\hline
\textbf{Likelihood functions} & \textbf{Logistic parameters} \\
\hline
$\mu_1$ & A : 0,
        K : 1,
        C : 1, 
        Q : 3,
        B : 20, 
        M: 2 \\
\hline
$\mu_2$ & A : 1,
        K : 0,
        C : 1, 
        Q : 50,
        B : 1.4, 
        M: 0.25 \\
\hline
\end{tabular}

\caption{The table shows the choice of criteria for the selection functions. Sets $A$ and $B$ are defined $\{(u, \mu(u))\mid u \in U\}$, in which $A$ is the selected fuzzy set and U is the set of all galaxies. The logistic parameters chosen for each likelihood function is specified.}  
\label{table:selections}
\end{table*}


\begin{table*}[t!]
\caption{Table of the candidate galaxies and their likelihood measures from different methods. The $^*$ and $^+$ next to the IDs show the candidates found in the primary and secondary samples reported in \cite{Merlin2019}, respectively. The $^*$ next to the redshift indicates those with available spectroscopic redshifts. The BBG, UVJ, and SED stand for likelihood based on the observed color, UVJ, and SED fitting, respectively. MIPS and X-ray columns correspond to the detection in the MIPS 24 $\mu$m and X-ray. $M_*$ is $\log{(M_s/M_\odot)}$, and Age is the age of the candidate galaxy in Gyrs. \textbf{The galaxies that show a detection in MIPS 24 $\mu$m and X-ray are excluded from the number/stellar mass densities measurements (i.e., Figure \ref{fig:n_density}).}}
\label{table:cand}
\centering
\begin{tabular}{|c|c|c|c|c|c|c|c|c|c|c|c|}
\hline
\textbf{Field} &
\textbf{ID} &
\textbf{RA} &
\textbf{DEC} &
\textbf{BBG} &
\textbf{UVJ} &
\textbf{SED} &
\textbf{$M_*$} &
\textbf{Age} &
\textbf{redshift} & 
\textbf{MIPS} &
\textbf{X-ray}\\
\hline
COSMOS & 3871 & 150.0839297 & 2.2235337 & 0.13 & 0.97 & 0.99 & 10.79 & 1.61 & 2.83 & False & False \\
COSMOS & 13639 & 150.0685942 & 2.3427211 & 0.35 & 0.98 & 0.95 & 10.86 & 1.46 & 2.93 & False & False \\
COSMOS & 14284 & 150.1060909 & 2.3510462 & 0.01 & 0.88 & 0.73 & 10.29 & 1.38 & 3.51 & False & False \\
COSMOS & 14403 & 150.109033 & 2.3523803 & 0.0 & 0.91 & 0.56 & 10.45 & 1.47 & 3.73 & False & False \\
COSMOS & 14528 & 150.1033018 & 2.3536424 & 0.0 & 0.46 & 0.35 & 10.63 & 1.11 & 3.12 & False & False \\ 
COSMOS & 14788 & 150.0543714 & 2.356883 & 0.07 & 0.33 & 0.78 & 11.2 & 1.11 & 4.02 & False & False \\
COSMOS & 16676$^*$ & 150.0614902 & 2.3786845 & 0.01 & 0.99 & 0.56 & 11.26 & 1.26 & 4.13 & False & True \\ 
COSMOS & 16726 & 150.102881 & 2.3794029 & 0.01 & 0.99 & 0.56 & 11.01 & 0.87 & 3.65 & True & False \\
COSMOS & 16948 & 150.0667157 & 2.3823608 & 0.04 & 0.91 & 0.43 & 10.8 & 0.87 & 3.70 & False & False \\ 
COSMOS & 18735 & 150.1557627 & 2.4044776 & 0.0 & 0.46 & 0.63 & 10.27 & 1.07 & 3.30 & False & False \\ 
COSMOS & 19502$^*$ & 150.1308597 & 2.4135984 & 0.22 & 0.86 & 0.58 & 10.78 & 0.84 & 3.87 & False & False \\ 
COSMOS & 21794 & 150.1707301 & 2.4442981 & 0.1 & 0.95 & 0.54 & 10.51 & 1.32 & 3.63 & False & False \\ 
COSMOS & 26635 & 150.0634065 & 2.5151526 & 0.13 & 0.92 & 0.61 & 10.43 & 1.08 & 3.24 & False & False \\ 
COSMOS & 26858 & 150.1648286 & 2.5190274 & 0.0 & 0.47 & 0.48 & 10.28 & 1.94 & 3.08 & True & False \\
COSMOS & 27856 & 150.0823166 & 2.5345944 & 0.18 & 0.95 & 0.67 & 10.79 & 0.87 & 4.08 & True & True \\ 
COSMOS & 31823 & 150.0597769 & 2.2810259 & 0.0 & 0.97 & 0.41 & 10.73 & 1.52 & 3.72 & True & False \\ 
COSMOS & 32689 & 150.2035188 & 2.3120491 & 0.0 & 0.55 & 0.73 & 10.62 & 1.47 & 3.64 & False & False \\ 
COSMOS & 33389 & 150.1758211 & 2.3387828 & 0.0 & 0.91 & 0.52 & 10.37 & 0.93 & 4.46 & False & False \\ 
COSMOS & 33927 & 150.0539523 & 2.3578909 & 0.0 & 0.46 & 0.4 & 10.04 & 1.26 & 3.25 & False & False \\ 
COSMOS & 33970 & 150.0799589 & 2.3592574 & 0.0 & 0.9 & 0.35 & 10.13 & 1.04 & 4.42 & False & False \\ 
COSMOS & 34944 & 150.0534942 & 2.3927052 & 0.0 & 0.42 & 0.58 & 10.19 & 1.66 & 3.53 & False & False \\ 
COSMOS & 35033 & 150.0530481 & 2.3964053 & 0.0 & 0.34 & 0.92 & 10.23 & 1.38 & 3.67 & False & False \\ 
COSMOS & 35098 & 150.0530811 & 2.3992994 & 0.0 & 0.46 & 0.84 & 10.14 & 1.54 & 3.19 & False & False \\ 
COSMOS & 35162 & 150.0536619 & 2.4022545 & 0.01 & 0.95 & 0.71 & 10.38 & 0.92 & 4.30 & False & False \\ 
COSMOS & 36674 & 150.1204151 & 2.4658729 & 0.0 & 0.48 & 0.84 & 10.18 & 1.86 & 3.19 & False & False \\ 
COSMOS & 37304 & 150.1030456 & 2.4952038 & 0.0 & 0.93 & 0.49 & 10.16 & 1.6 & 3.57 & False & False \\ 
COSMOS & 38150 & 150.0825841 & 2.5312043 & 0.0 & 0.99 & 0.35 & 10.2 & 0.99 & 4.29 & False & False \\
\hline \hline

 EGS & 8 & 215.300542 & 53.051323 & 0.0 & 0.93 & 0.4 & 11.06 & 1.59 & 3.76 & True & False \\ 
 EGS & 2922 & 214.932074 & 52.818233 & 0.0 & 0.99 & 0.46 & 10.43 & 1.39 & 3.29 & False & False \\ 
 EGS & 6162 & 215.041352 & 52.914091 & 0.0 & 0.96 & 0.39 & 10.87 & 1.58 & 3.21 & True & False \\ 
 EGS & 6498 & 215.065871 & 52.932958 & 0.0 & 0.76 & 0.55 & 10.53 & 1.72 & 3.45 & False & False \\ 
 EGS & 14727$^*$ & 214.895659 & 52.856515 & 0.0 & 0.8 & 0.57 & 10.98 & 1.2 & 3.05 & True & False \\ 
 EGS & 16431 & 215.191335 & 53.074718 & 0.0 & 0.96 & 0.5 & 10.45 & 1.09 & 3.49 & False & False \\ 
 EGS & 21158 & 214.746219 & 52.783393 & 0.0 & 0.98 & 0.44 & 11.0 & 1.34 & 3.90 & True & False \\
 EGS & 21351$^*$ & 214.673655 & 52.732542 & 0.3 & 0.96 & 0.5 & 10.59 & 0.88 & 3.61 & False & False \\ 
 EGS & 22706 & 215.122517 & 53.058015 & 0.0 & 0.98 & 0.51 & 10.5 & 1.22 & 3.25 & False & False \\ 
 EGS & 23036$^*$ & 214.879114 & 52.88807 & 0.0 & 0.73 & 0.46 & 10.22 & 0.98 & 3.56 & False & False \\ 
 EGS & 23572 & 215.144538 & 53.078392 & 0.0 & 0.98 & 0.44 & 10.07 & 1.64 & 3.15 & False & True \\ 
 EGS & 24177$^*$ & 214.866081 & 52.884232 & 0.0 & 0.92 & 0.58 & 10.98 & 1.78 & 3.42 & True & False \\ 
 EGS & 24356$^+$ & 214.620084 & 52.70959 & 0.06 & 0.92 & 0.47 & 10.66 & 1.53 & 3.43 & False & False \\ 
 EGS & 24948 & 214.767294 & 52.81771 & 0.0 & 0.75 & 0.52 & 10.38 & 1.38 & 3.44 & False & False \\ 
 EGS & 25724$^*$ & 214.997776 & 52.986129 & 0.0 & 0.96 & 0.5 & 10.6 & 1.43 & 3.80 & False & False \\ 
 EGS & 27491$^+$ & 214.617755 & 52.724101 & 0.56 & 0.92 & 0.46 & 10.55 & 1.18 & 3.34 & False & False \\ 
 EGS & 29547$^*$ & 214.695306 & 52.796871 & 0.16 & 0.97 & 0.51 & 10.54 & 1.59 & 3.15 & False & False \\ 
 EGS & 30198 & 214.966237 & 52.983055 & 0.0 & 0.49 & 0.42 & 10.03 & 1.26 & 3.01 & False & False \\ 
 EGS & 30619 & 214.981814 & 52.991238 & 0.0 & 0.94 & 0.57 & 10.56 & 1.95 & 3.07 & False & False \\ 
 EGS & 32592 & 215.080479 & 52.921568 & 0.0 & 0.97 & 0.53 & 10.2 & 0.97 & 4.28 & False & False \\ 
 EGS & 33316 & 215.252837 & 53.055545 & 0.0 & 0.96 & 0.54 & 10.47 & 1.04 & 3.70 & False & False \\ 
 EGS & 35080 & 214.922243 & 52.854748 & 0.0 & 0.79 & 0.59 & 10.53 & 0.9 & 4.00 & False & False \\ 
 EGS & 35459 & 215.231097 & 53.079181 & 0.0 & 0.88 & 0.32 & 10.0 & 1.31 & 3.95 & False & False \\ 
 EGS & 36375 & 214.778097 & 52.774151 & 0.0 & 0.46 & 0.34 & 10.51 & 1.6 & 3.27 & False & False \\ 
 EGS & 38679 & 214.692609 & 52.753274 & 0.0 & 0.96 & 0.65 & 10.92 & 1.36 & 3.91 & False & False \\ 
 EGS & 41385 & 214.701834 & 52.812333 & 0.0 & 0.35 & 0.33 & 10.37 & 0.96 & 3.63 & True & False \\
 \hline \hline 
 GOODSN & 599 & 189.22477876 & 62.12097508 & 0.0 & 0.98 & 0.37 & 10.7 & 1.3 & 3.77 & False & False \\ 
GOODSN & 3225 & 188.99068993 & 62.16051335 & 0.13 & 0.45 & 0.94 & 10.04 & 1.98 & 3.18 & False & False \\
GOODSN & 4004$^*$ & 189.26573822 & 62.16839485 & 0.54 & 0.77 & 0.63 & 10.31 & 1.03 & 3.81 & False & False \\ 
GOODSN & 4572 & 189.32906181 & 62.17385428 & 0.56 & 0.63 & 0.53 & 10.37 & 1.0 & 4.15 & True & True \\ 
\hline
\end{tabular}
\end{table*}

\begin{table*}[t!]
\centering
\resizebox{\columnwidth}{!}{
\begin{tabular}{|c|c|c|c|c|c|c|c|c|c|c|c|}
\hline
\textbf{Field} &
\textbf{ID} &
\textbf{RA} &
\textbf{DEC} &
\textbf{BBG} &
\textbf{UVJ} &
\textbf{SED} &
\textbf{$M_*$} &
\textbf{Age} &
\textbf{redshift} & 
\textbf{MIPS} &
\textbf{X-ray}\\
\hline
GOODSN & 4691$^+$ & 189.10990126 & 62.17519651 & 0.88 & 0.99 & 0.48 & 10.7 & 1.64 & 3.18 & False & False \\
GOODSN & 5059$^*$ & 189.1623096 & 62.17823976 & 0.01 & 0.98 & 0.51 & 10.79 & 1.08 & 3.69 & False & False \\ 
GOODSN & 5744$^*$ & 189.10012474 & 62.18362694 & 0.05 & 0.98 & 0.51 & 10.39 & 1.11 & 3.46 & False & False \\ 
GOODSN & 6989 & 189.22148283 & 62.19241118 & 0.48 & 0.98 & 0.98 & 10.22 & 1.65 & 2.80 & False & False \\ 
GOODSN & 8074 & 189.15625504 & 62.19909581 & 0.48 & 0.93 & 1.0 & 11.86 & 0.63 & 5.11 & True & False \\ 
GOODSN & 8109 & 189.26660705 & 62.19930947 & 0.01 & 0.47 & 0.89 & 10.63 & 1.68 & 3.41$^*$ & True & False \\ 
GOODSN & 9083 & 189.33429106 & 62.20610777 & 0.69 & 0.58 & 0.2 & 10.11 & 1.53 & 3.62 & True & False \\ 
GOODSN & 9545 & 188.98263204 & 62.20890348 & 0.16 & 0.34 & 0.36 & 10.41 & 1.42 & 4.27 & False & False \\ 
GOODSN & 24582 & 189.40615811 & 62.34785252 & 0.5 & 0.35 & 0.0 & 9.2 & 1.95 & 2.87 & False & False \\ 
GOODSN & 27305 & 189.21770656 & 62.31096041 & 0.0 & 0.98 & 0.51 & 10.18 & 2.03 & 3.01 & False & False \\
\hline
\hline
GOODSS & 2032 & 53.2293854 & -27.8977318 & 0.0 & 0.96 & 0.48 & 10.07 & 1.09 & 3.08 & False & False \\ 
GOODSS & 2717$^*$ & 53.1893272 & -27.8884506 & 0.0 & 0.99 & 0.48 & 10.92 & 1.48 & 3.03 & False & False \\ 
GOODSS & 2782$^*$ & 53.0835724 & -27.8875294 & 0.75 & 0.92 & 0.49 & 10.48 & 1.53 & 3.58 & False & False \\ 
GOODSS & 3912$^*$ & 53.0622215 & -27.8749809 & 0.35 & 0.89 & 0.04 & 10.17 & 1.34 & 3.90 & False & False \\ 
GOODSS & 4503$^+$ & 53.1132774 & -27.869875 & 0.0 & 0.86 & 0.42 & 10.91 & 1.29 & 3.59 & False & False \\
GOODSS & 4821 & 53.0825539 & -27.866745 & 0.04 & 0.75 & 0.51 & 10.24 & 1.5 & 3.10 & False & True \\ 
GOODSS & 5479 & 53.0784645 & -27.8598576 & 0.83 & 0.71 & 0.26 & 11.07 & 1.15 & 3.66$^*$ & True & False \\ 
GOODSS & 6131 & 53.0916061 & -27.8533421 & 0.47 & 0.96 & 1.0 & 11.74 & 1.1 & 5.06 & True & False \\ 
GOODSS & 6235 & 53.1199837 & -27.8519554 & 0.36 & 0.32 & 0.0 & 10.19 & 1.66 & 3.50 & False & False \\
GOODSS & 7526$^+$ & 53.0786781 & -27.8395462 & 0.25 & 0.97 & 0.44 & 10.14 & 1.32 & 3.32 & False & False \\ 
GOODSS & 8785$^*$ & 53.0818481 & -27.8287373 & 0.61 & 0.98 & 0.48 & 10.23 & 1.47 & 3.85 & False & False \\ 
GOODSS & 9209$^*$ & 53.1081772 & -27.8251228 & 0.31 & 0.63 & 0.49 & 10.65 & 1.2 & 4.49 & False & False \\ 
GOODSS & 12178$^+$ & 53.0392838 & -27.7993088 & 0.87 & 0.79 & 0.04 & 10.36 & 1.62 & 3.29 & False & True \\ 
GOODSS & 12407 & 53.2218704 & -27.7976608 & 0.0 & 0.97 & 0.34 & 11.06 & 1.03 & 4.24 & True & False \\ 
GOODSS & 13299 & 53.2072105 & -27.7913475 & 0.53 & 0.43 & 0.0 & 10.22 & 0.91 & 4.05 & False & False \\ 
GOODSS & 16671 & 53.1901817 & -27.7691402 & 0.51 & 0.98 & 0.77 & 10.46 & 1.59 & 2.87 & False & False \\ 
GOODSS & 17258 & 53.1411247 & -27.7643566 & 0.31 & 0.32 & 0.0 & 10.4 & 1.31 & 4.52 & False & False \\ 
GOODSS & 17749$^*$ & 53.1968956 & -27.7604523 & 0.99 & 0.99 & 0.49 & 10.69 & 1.69 & 3.70 & True & False \\
GOODSS & 18180$^*$ & 53.1812248 & -27.756422 & 0.98 & 0.96 & 0.46 & 10.62 & 1.01 & 3.65 & False & False \\ 
GOODSS & 19883$^*$ & 53.0106544 & -27.7416039 & 0.0 & 0.97 & 0.42 & 10.87 & 0.97 & 3.57 & True & True \\ 
GOODSS & 20111 & 53.0944252 & -27.7392502 & 0.0 & 0.32 & 0.56 & 10.82 & 1.08 & 3.13 & True & False \\ 
GOODSS & 22085$^*$ & 53.0738754 & -27.7221718 & 0.0 & 0.92 & 0.48 & 10.36 & 1.17 & 3.47 & False & False \\ 
GOODSS & 32527 & 53.0305367 & -27.7529354 & 0.0 & 0.96 & 0.35 & 10.48 & 0.83 & 4.28 & True & False \\
\hline \hline
UDS & 164 & 34.3170586 & -5.2759099 & 0.0 & 0.44 & 0.38 & 10.46 & 0.92 & 3.53 & False & False \\ 
UDS & 416 & 34.5199471 & -5.2747688 & 0.0 & 0.98 & 0.46 & 11.39 & 1.33 & 3.38 & True & False \\ 
UDS & 635 & 34.5062943 & -5.273128 & 0.0 & 0.99 & 0.53 & 11.0 & 0.92 & 4.14 & False & False \\ 
UDS & 918 & 34.2661095 & -5.2721262 & 0.0 & 0.77 & 0.44 & 10.88 & 1.09 & 3.17 & True & False \\ 
UDS & 1244$^*$ & 34.2894669 & -5.269805 & 0.32 & 0.89 & 0.01 & 10.77 & 0.99 & 3.79 & False & False \\ 
UDS & 1408 & 34.5122528 & -5.2688479 & 0.0 & 0.96 & 0.49 & 10.77 & 0.98 & 4.16 & False & False \\ 
UDS & 2571$^*$ & 34.290432 & -5.2620749 & 0.7 & 0.8 & 0.08 & 10.52 & 1.39 & 3.70 & False & False \\ 
UDS & 3752 & 34.5192871 & -5.2553701 & 0.0 & 0.73 & 0.52 & 10.1 & 1.5 & 3.11 & False & False \\ 
UDS & 4319 & 34.4653702 & -5.2520308 & 0.23 & 0.99 & 0.51 & 11.69 & 0.89 & 4.48 & True & False \\ 
UDS & 4332$^+$& 34.4656906 & -5.2519188 & 0.01 & 0.93 & 0.57 & 10.98 & 1.7 & 3.18 & True & False \\ 
UDS & 5256 & 34.2441864 & -5.2458172 & 0.0 & 0.34 & 0.89 & 10.89 & 1.86 & 3.31 & False & False \\ 
UDS & 6218 & 34.3409538 & -5.2405558 & 0.61 & 0.76 & 0.01 & 10.89 & 0.93 & 4.07 & False & False \\ 
UDS & 7520$^*$ & 34.2558746 & -5.23383 & 0.57 & 0.96 & 0.46 & 11.1 & 1.48 & 3.17 & True & False \\ 
UDS & 7779$^+$ & 34.2588844 & -5.2323041 & 0.54 & 0.48 & 0.29 & 10.66 & 1.75 & 3.14 & False & False \\ 
UDS & 8682$^+$ & 34.2937317 & -5.2269621 & 0.82 & 0.92 & 0.26 & 10.44 & 1.58 & 3.46 & False & False \\
UDS & 13988 & 34.3857307 & -5.1989851 & 0.36 & 0.05 & 0.63 & 11.45 & 1.95 & 3.03 & True & False \\ 
UDS & 15748 & 34.5302429 & -5.1890779 & 0.0 & 0.46 & 0.36 & 11.26 & 2.05 & 3.08 & True & False \\ 
UDS & 17344 & 34.3231277 & -5.179821 & 0.35 & 0.16 & 0.44 & 10.88 & 1.87 & 3.03 & True & False \\ 
UDS & 17790 & 34.5422859 & -5.1774998 & 0.34 & 0.33 & 0.0 & 10.54 & 1.58 & 3.31 & True & False \\ 
UDS & 18672 & 34.5668411 & -5.1726952 & 0.92 & 0.51 & 0.0 & 10.32 & 1.32 & 3.79 & False & False \\ 
UDS & 19849 & 34.3381882 & -5.1661878 & 0.07 & 0.98 & 0.41 & 10.41 & 1.66 & 3.53 & True & False \\ 
UDS & 20843$^*$ & 34.4961014 & -5.161037 & 0.64 & 0.65 & 0.19 & 10.76 & 0.88 & 3.74 & False & False \\ 
UDS & 22354 & 34.4276466 & -5.1524191 & 0.72 & 0.38 & 0.03 & 10.93 & 1.9 & 3.15 & False & False \\ 
UDS & 23628$^*$ & 34.2425995 & -5.1430721 & 0.97 & 0.63 & 0.21 & 10.73 & 0.83 & 4.25 & False & False \\ 
UDS & 24501 & 34.5228386 & -5.1288252 & 0.0 & 0.82 & 0.34 & 10.43 & 1.55 & 3.40 & False & False \\ 
UDS & 24734 & 34.5229988 & -5.1295991 & 0.0 & 0.88 & 0.49 & 10.36 & 1.55 & 3.47 & False & False \\ 
UDS & 25688$^*$ & 34.5265884 & -5.1360388 & 0.25 & 0.99 & 0.53 & 11.21 & 1.33 & 3.08 & False & False \\ 
UDS & 25893$^*$ & 34.3996353 & -5.1363459 & 0.08 & 0.99 & 0.55 & 11.05 & 1.24 & 4.49 & False & False \\ 
UDS & 32698 & 34.5237198 & -5.1804299 & 0.0 & 0.36 & 0.61 & 10.99 & 1.29 & 4.30 & True & False \\ 
UDS & 35635 & 34.3170319 & -5.127574 & 0.0 & 0.93 & 0.36 & 10.58 & 1.36 & 3.72 & False & True \\
\hline
\end{tabular}
}
\end{table*}

\end{document}